\documentclass[sigconf]{acmart}

\pdfoutput=1

\AtBeginDocument{%
  \providecommand\BibTeX{{%
    Bib\TeX}}}

\copyrightyear{2026}
\acmYear{2026}
\setcopyright{cc}
\setcctype{by}
\acmConference[ICSE '26]{2026 IEEE/ACM 48th International Conference on Software Engineering}{April 12--18, 2026}{Rio de Janeiro, Brazil}
\acmBooktitle{2026 IEEE/ACM 48th International Conference on Software Engineering (ICSE '26), April 12--18, 2026, Rio de Janeiro, Brazil}
\acmPrice{}
\acmDOI{10.1145/3744916.3773108}
\acmISBN{979-8-4007-2025-3/2026/04}

\usepackage{hyperref}
\usepackage{graphicx}
\usepackage{setspace}
\usepackage{graphicx}
\usepackage{setspace}
\usepackage{booktabs}
\usepackage{caption}

\usepackage{tabularx}
\usepackage{amsmath}

\usepackage{multirow}
\usepackage{subcaption}
\usepackage{xspace}
\usepackage{verbatim}
\usepackage{color}
\usepackage{url}
\usepackage{listings}
\usepackage{algorithm}
\usepackage{mathtools}
\usepackage{enumitem}
\usepackage{textcomp}
\usepackage{cleveref} 
\usepackage{float}
\usepackage{colortbl}
\usepackage{framed}

\newcommand{\mybox}[1]{%
  \par\vspace{5pt}\noindent%
  {%
    \setlength{\fboxsep}{5pt}
    \colorbox{gray!15}{%
      \parbox{\dimexpr\linewidth-2\fboxsep\relax}{
        \emph{#1}
      }%
    }%
  }%
  \par\vspace{5pt}%
}%

\def\BibTeX{{\rm B\kern-.05em{\sc i\kern-.025em b}\kern-.08emT\kern-.1667em\lower.7ex\hbox{E}\kern-.125emX}}

\newcommand{\remove}[1]{}
\newcommand{\git}[1]{\textit{GitHub}}

\newcommand{\cmr}[1]{{#1}}

\newcommand{\modify}[1]{#1}
\newcommand{\gr}[1]{#1}

\newdimen{\algindent}
\definecolor{codegreen}{rgb}{0,0.6,0}
\definecolor{codegray}{rgb}{0.5,0.5,0.5}
\definecolor{uncoveredgray}{rgb}{0.8,0.8,0.8}
\definecolor{codepurple}{rgb}{0.58,0,0.82}
\definecolor{backcolour}{rgb}{0.99,0.80,0.87}

\newcommand{\parabf}[1]{\noindent\textbf{#1}}

\newcommand{\rustbench}{Rust-SWE-bench}

\newcommand{\swebench}{SWE-bench}
\newcommand{\swebenchverified}{SWE-bench Verified}
\newcommand{\repoCNT}{87}
\newcommand{\selectedRepoCNT}{34}
\newcommand{\candidateTaskCNT}{1,040}
\newcommand{\prCNT}{80,000}
\newcommand{\githubstar}{1,000}
\newcommand{\approachUQResolve}{46}
\newcommand{\agentless}{\textsc{Agentless}}
\newcommand{\sweagent}{\textsc{SWE-agent}}
\newcommand{\openhands}{\textsc{OpenHands}} 
\newcommand{\autocode}{\textsc{AutoCodeRover}} 

\newcommand{\claudesonnet}{Claude-Sonnet-3.7}
\newcommand{\gpto}{GPT-4o}
\newcommand{\omini}{OpenAI o4-mini}
\newcommand{\qwen}{Qwen3}

\newcommand{\rustagent}{\textsc{RustForger}}

\begin{document}

\title{Evaluating and Improving Automated Repository-Level Rust Issue Resolution with LLM-based Agents}

\author{Jiahong Xiang\textsuperscript{\textdagger{}}}
\affiliation{
    \institution{Research Institute of Trustworthy \\ Autonomous Systems, Southern University of Science and Technology}
  \city{Shenzhen}
  \country{China}}
\email{xiangjh2022@mail.sustech.edu.cn}

\author{Wenxiao He}
\affiliation{
  \institution{Southern University of Science and Technology}
  \city{Shenzhen}
  \country{China}}
\email{12110818@mail.sustech.edu.cn}

\author{Xihua Wang}
\affiliation{
  \institution{Southern University of Science and Technology}
  \city{Shenzhen}
  \country{China}}
\email{12213006@mail.sustech.edu.cn}

\author{Hongliang Tian}
\affiliation{%
  \institution{Ant Group}
  \city{Hangzhou}
  \country{China}}
\email{tate.thl@antgroup.com}

\author{Yuqun Zhang\textsuperscript{\textdagger{}}*}
\affiliation{%
  \institution{Research Institute of Trustworthy \\ Autonomous Systems, Southern University of Science and Technology}
  \city{Shenzhen}
  \country{China}}
\email{zhangyq@sustech.edu.cn}

\thanks{\textsuperscript{*}Yuqun Zhang is the corresponding author.}
\thanks{\textsuperscript{\textdagger{}}These authors are also affiliated with the Department of Computer Science and Engineering, Southern University of Science and Technology, Shenzhen, China.}

\begin{abstract}

The Rust programming language presents a steep learning curve and significant coding challenges, making the automation of issue resolution essential for its broader adoption. Recently, LLM-powered code agents have shown remarkable success in resolving complex software engineering tasks, yet their application to Rust has been limited by the absence of a large-scale, repository-level benchmark. To bridge this gap, we introduce \rustbench{}, a benchmark comprising 500 real-world, repository-level software engineering tasks from \selectedRepoCNT{} diverse and popular Rust repositories. We then perform a comprehensive study on \rustbench{} with four representative agents and four state-of-the-art LLMs to establish a foundational understanding of their capabilities and limitations in the Rust ecosystem.

Our extensive study reveals that while ReAct-style agents are promising, i.e., resolving up to 21.2\% of issues, they are limited by two primary challenges: comprehending repository-wide code structure and complying with Rust's strict type and trait semantics. We also find that issue reproduction \modify{is} rather critical for task resolution. Inspired by these findings, we propose \rustagent{}, a novel agentic approach that integrates an automated test environment setup with a Rust metaprogramming-driven dynamic tracing strategy to facilitate reliable issue reproduction and \modify{dynamic} analysis. The evaluation shows that \rustagent{} \modify{using \claudesonnet{}} significantly outperforms all baselines, resolving 28.6\% of tasks on \rustbench{}, i.e., a 34.9\% improvement over the strongest baseline, and\modify{, in aggregate,} uniquely solves \approachUQResolve{} tasks that no other agent could solve \modify{across all adopted advanced LLMs}.

\end{abstract}



\maketitle


\section{Introduction}
The Rust programming language has become increasingly popular by offering a compelling combination of performance and safety~\cite{Shanker2018Concurrency, RustTeam2019, Chua2017MemorySafety}. Through its strict type system and ownership model, it provides compile-time guarantees against memory faults and data races~\cite{RustLang2021}, which has led to its wide adoption in critical domains such as operating systems~\cite{RedoxOS, Li2024RustForLinux,rustylinux}, cloud services~\cite{Agache2020Firecracker}, and web browsers~\cite{Servo2019, Jansens2023Chromium, Mozilla2021Foundation}. 
However, its core strengths, i.e., memory and thread safety, also cause its steep learning curve and coding difficulty, posing significant challenges for developers, with 83\% of them finding it difficult to use~\cite{RustSurvey2021} and 42\%  worrying about the long "time to productivity"~\cite{Fulton2021Benefits}. Therefore, automating the resolution of the issues related to development in Rust is rather essential for unlocking Rust's full potential and extending its adoption.

Recently, Large Language Models (LLMs) have demonstrated powers in addressing a wide range of software engineering challenges
~\cite{swebench,specrover,libro,tan2024llm4decompile,tan2025decompile,tan2025sk2decompile,tan2024prompt,xiang2024far,gao2025oasis,liang2025grammar}. 
Specifically, LLM code agents, which leverage LLMs' coding abilities through tool adoption, command execution, environmental feedback, and action planning, have demonstrated superior performance in code-related tasks~\cite{sweagent, anthropic_claude4_2025, openhands} and aided developers in coding efficiency~\cite{barke2023grounded, tang2024study}. Their strengths are typically exemplified by their performance on popular benchmarks such as \swebenchverified{}~\cite{swebench}, a version of \swebench{} crafted by OpenAI~\cite{OpenAI2024SWEBench} for resolving real-world Python issues. In this benchmark, an agent, given only an issue description and a full repository codebase, is expected to autonomously generate a single corrective patch to resolve the issue and pass all tests in one attempt. This task is highly demanding, requiring a suite of capabilities including repository-level code understanding, targeted search, issue-reproducing test generation, and precise code editing. 
Surprisingly, the resolution rate on this benchmark has been improving dramatically, from a mere 1.2\% achieved by approaches like RAG with SWE-Llama 13B in October 2023~\cite{swebench} to a recent high  80.2\% achieved by agents like Claude-Tools powered by the Claude 4 Opus model~\cite{anthropic_claude4_2025}.

The demand \gr{for} automating Rust issue resolution and the demonstrated power of advanced LLMs and code agents altogether raise a compelling question: \textit{can agentic approaches effectively automate the resolution of real-world Rust issues?} To answer this question, it is essential for a large-scale, repository-level benchmark for real-world \modify{software engineering} tasks for Rust and a comprehensive study on how advanced LLMs and agents perform on such tasks upon the benchmark. However, existing benchmarks for Rust are deficient in two key aspects. 
First, while existing Rust benchmarks focus on granular tasks such as C-to-Rust transpilation~\cite{ou2024repository, athiwaratkun2022multi,khan2023xcodeeval}, function-level code synthesis~\cite{octopack,cassano2023multipl}, or specific CVE analysis~\cite{ni2024towards,xu2021memory,qin2020understanding,zheng2023closer}, they are limited in involving general, repository-level software engineering problems.
Second, existing large-scale SWE benchmarks predominantly feature Python and Java~\cite{swebench, repocoder, repobench}. The few that incorporate Rust only offer a smaller corpus of tasks (e.g., 43 in SWE-bench Multilingual~\cite{khandpur2025multilingual} and 239 in Multi-SWE-bench~\cite{mswe} \modify{compared with 500 Python issues in \swebenchverified{}~\cite{OpenAI2024SWEBench}}). 
Consequently, the absence of a benchmark that is both large-scale and dedicated to real-world Rust software engineering tasks prevents us from comprehensively assessing the effectiveness of the LLMs and agents on resolving practical, repository-level Rust issues.

To bridge this gap, we introduce \rustbench{}, a large-scale benchmark that contains 500 real-world, issue-resolving \modify{repository-level} Rust tasks from \selectedRepoCNT{} popular repositories. Specifically, we adopt the well-established \swebench{} workflow~\cite{swebench} and  adhere to the verification protocol established by the popular \swebenchverified{}~\cite{OpenAI2024SWEBench} to construct it. Our data collection process begins by sourcing pull requests (PRs) from approximately \repoCNT{} prominent open-source Rust repositories selected based on their popularity on GitHub (> 1,000 stars). 
We then identify candidate tasks with PRs that are both merged and linked to a resolved GitHub issue and that contain modifications to test files. Finally, to ensure that each task represents a verifiable fix, we validate whether each patch induces a "fail-to-pass" transition (i.e., at least one test fails before the patch and passes after the patch is applied), yielding a candidate set of \candidateTaskCNT{} Rust \modify{software engineering} tasks. To mitigate issues like underspecified problem descriptions~\cite{agentless, OpenAI2024SWEBench} which can lead to the underestimation of LLM capabilities and inefficient use of computational resources, we perform a final manual inspection of task candidates, following OpenAI's construction principles~\cite{OpenAI2024SWEBench}. Eventually, we obtain 500 high-quality \modify{software engineering} tasks for Rust from \selectedRepoCNT{} diverse GitHub repositories.

To assess the capabilities of advanced LLMs and agents on this new benchmark, we conduct a comprehensive empirical study involving four representative agentic approaches (i.e., \sweagent{}~\cite{sweagent}, \openhands{}+CodeAct v2.1~\cite{openhands}, \agentless{}~\cite{agentless}, and \autocode{} v2.0~\cite{specrover}) and four state-of-the-art LLMs (i.e., \claudesonnet{}~\cite{anthropic2024claude3_7}, \gpto{}~\cite{openai2024gpt4o}, \omini{}~\cite{openai2025o3o4mini}, and \qwen{}~\cite{Qwen_Pricing}). Specifically, these studied agents are selected for their general-purpose design, i.e., having demonstrated capabilities across various programming languages~\cite{mswe, swepoly, khandpur2025multilingual}.
The evaluation results demonstrate that the adoption of a ReAct-style~\cite{react} (`thought-act-observe' loop) architecture excels in solving tasks on \rustbench{}. The associated agent \openhands{} with \claudesonnet{}~\cite{anthropic2024claude3_7} resolves a substantial 21.2\% of real-world \modify{repository-level} Rust issues, for which the manual resolution process averages 126 days and approximately 5.5 rounds of discussion. Moreover, we find that top-performing agents distinguish themselves by their ability to generate large, complex patches, e.g., the leading agent \openhands{} resolves significantly more issues than \agentless{} (105 vs. 57) when the required patch size exceeds 150 lines. 
We also find the compilation errors stem from failures to model repository-wide code structure and to comply with Rust’s strict type and trait semantics.
Interestingly, we observe that issue reproduction is critical for Rust issue resolution, e.g., the remaining 44.5\% of tasks fail at the reproduction stage and thus could not be successfully resolved even under the top-performing agent-model configuration.

To mitigate the issues of repository-wide code analysis and \modify{issue} reproduction in our study findings, we further introduce \rustagent{}, a novel agentic approach through automated test environment setup coupled with a cross-project dynamic tracing mechanism.
Specifically, \rustagent{} first constructs an isolated testing workspace, managing Cargo dependencies~\cite{rust-cargo-book}  to create a reliable and sandboxed environment for issue reproduction. 
Within this controlled environment, the agent can invoke a novel \texttt{Trace} command for cross-project dynamic analysis. \modify{This command automates the entire dynamic analysis workflow} by leveraging Rust's metaprogramming capabilities—specifically procedural macros—to automatically instrument target functions through Abstract Syntax Tree (AST) modification. Such a decoupled strategy enables the agent to capture precise runtime control-flow information by executing tests from the clean workspace, thereby bypassing the complex build systems of the original project.
Accordingly, \rustagent{} equips the agent with the necessary runtime insights to more effectively diagnose and resolve complex real-world Rust issues. The evaluation results show \rustagent{} solves \approachUQResolve{} tasks that cannot be solved by any studied agent across all LLMs; its pairing with \claudesonnet{} resolves 28.6\% of tasks—a 34.9\% improvement over the strongest baseline.

In summary, this paper makes the following contributions:

\begin{itemize}[leftmargin=*]
    \item \textbf{Benchmark.} We construct \rustbench{}, the first large-scale, repository-level benchmark dedicated to real-world Rust software engineering issues. Comprising 500 high-quality, verified tasks from \selectedRepoCNT{} popular repositories, \rustbench{} addresses a critical gap in the field and enables systematic evaluation and future research on automated Rust development.

    \item \textbf{Study.} We conduct a comprehensive empirical study on \rustbench{} with four representative agents and four state-of-the-art LLMs. 
    Our analysis reveals that while ReAct-style agents show promising results, they are limited by two key challenges: comprehending repository-wide code structure and semantics, and \modify{successfully reproducing issues.}

    \item \textbf{\rustagent{}.} We propose \rustagent{}, a novel agentic framework that addresses identified challenges by integrating an automated testing workspace with a cross-project dynamic tracing mechanism, powered by the distinctive programming features of Rust. Our evaluation shows that \rustagent{} resolves 28.6\% of tasks in the \rustbench{} benchmark using \claudesonnet{}, significantly outperforming other baselines, and uniquely resolves \approachUQResolve{} tasks across all adopted LLMs.
\end{itemize}


\section{Background \& Related Work}

\subsection{Rust Programming Language}

Rust offers a modern solution for high-performance and secure software that provides strong compile-time guarantees against memory faults and data races without the overhead of a garbage collector. This combination of safety and performance has spurred its adoption in critical domains, including operating systems such as Redox~\cite{redox} and Tock~\cite{tockos}, web browsers such as Servo~\cite{desosa2019servo}, and cloud infrastructure like Firecracker~\cite{Agache2020Firecracker}.
\modify{Notably, Rust now constitutes 21\% of all new native code in the Android OS codebase~\cite{Li2022}.}

While Rust's safety guarantees are enforced by its unique type system (built on ownership, borrowing, and lifetimes), such a system is notoriously difficult for newcomers to master, creating a steep learning curve and a significant barrier for adoption~\cite{RustSurvey2021, rustcasestudyrust2023}. To address this high development overhead, researchers and developers have proposed various approaches. RustAssistant~\cite{deligiannis2023fixing} leverages LLMs to help developers automatically fix compilation errors. The LLM-driven Syzygy~\cite{shetty2024syzygy} aims to automate the migration of legacy C codebases to Rust. Concrat~\cite{hong2023concrat} analyzes and replaces unsafe C lock APIs in concurrent programs with provably safe Rust equivalents. Bronze~\cite{coblenz2022garbage} introduces an optional garbage collector to reduce the difficulty of complex memory management. PanicKiller~\cite{ni2024towards} provides the first automated tool for fixing panic bugs in real-world Rust programs. However, despite these specialized solutions, a general-purpose agent capable of resolving diverse, real-world Rust issues remains an open challenge.

\subsection{Large Language Model-based Agents}

\begin{figure}[!htb]
    \centering
    \includegraphics[width=0.89\columnwidth]{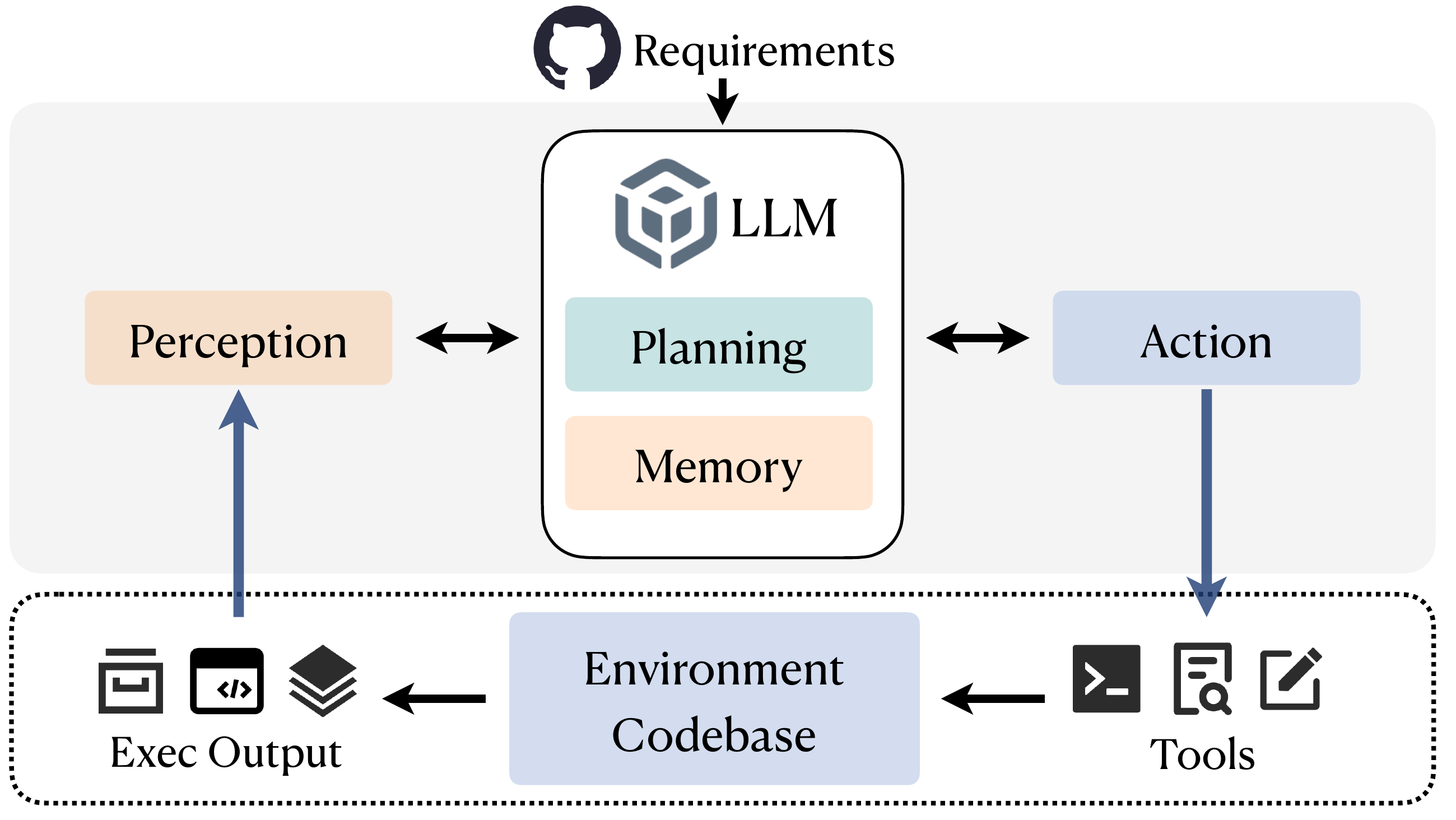}
    \caption{Basic Framework of LLM-based Agents}
    \label{fig:llm}
\end{figure}

LLM-based agents refer to autonomous systems that use an LLM as a central controller to perceive and act upon an environment to achieve specific goals~\cite{jin2025llmsllmbasedagentssoftware, wang2024survey}. \modify{These agents operate in an iterative cycle, leveraging core components: planning for decomposing complex tasks, memory for learning from historical observations~\cite{zhang2024survey}, and perception for processing environmental feedback, as shown in Figure~\ref{fig:llm}.} Crucially, their action component enables them to utilize external tools, allowing them to interact with and modify their environment in ways far beyond simple text generation~\cite{wang2024survey}. 

Researchers have attempted to use agents to automatically resolve real-world, repository-level software engineering tasks~\cite{swebench, khandpur2025multilingual, mswe, swepoly, coderujb, codereval, complexcodeeval}. Specifically, SWE-bench~\cite{swebench} has gained significant attention since its release, with numerous agents being evaluated on this benchmark. In this benchmark, an agent, given only an issue description and the repository codebase, is expected to autonomously generate a single corrective patch to resolve the issue and pass all tests in one attempt. This task is highly demanding, requiring a suite of capabilities including repository-level code understanding, targeted search, \modify{issue-reproducing} test generation, and precise code editing. 
\modify{Notably, reproducing an issue—whether a bug, a missing feature, or an API behavior—is a critical step for capturing dynamic execution information, which in turn facilitates more precise issue-related code localization and targeted patch generation~\cite{le2015information} and serves as a foundational stage in recent advanced agents~\cite{agentless, specrover, sweagent, openhands, coder}.} 

To resolve the tasks mentioned above, a variety of paradigms have been proposed. ReAct~\cite{react} enables an agent to synergize reasoning and acting by interleaving a `thought-act-observe' loop. For instance, \sweagent{}~\cite{sweagent} implements the ReAct loop with an Agent-Computer Interface (ACI) by equipping LLMs with high-level tools for file searching, editing, and navigation, instead of raw terminal interaction. Similarly, the \openhands{} agent~\cite{openhands}, built upon the ReAct-style CodeAct architecture~\cite{codeact}, operates as a generalist platform where agents perform tasks by executing code or conversing with humans for clarification. 
Another group of agents employ more structured, multi-stage workflows. For instance, \autocode{} v2.0~\cite{specrover} utilizes a pipeline of specialized agents for iterative context retrieval and specification inference to guide patch generation. \agentless{}~\cite{agentless} demonstrates the efficacy of a simpler, non-iterative workflow (e.g., localize, repair, and validate) that deliberately limits the LLM’s autonomy. While these agents show promising results on Python-centric benchmarks like \swebenchverified{} (e.g., \openhands{} at 70.4\% and \agentless{} at 50.8\%), their effectiveness on real-world Rust issues remains underexplored.

\subsection{SWE-related Benchmarks} 
\label{ref:swe-related-benchs}

\swebench{}~\cite{swebench} is a prominent benchmark for evaluating end-to-end software maintenance capabilities, comprising 2,294 real-world tasks derived from GitHub issues across 12 popular Python repositories. \modify{Specifically, the objective for each task is to automatically resolve the corresponding issue and submit a patch.} The construction of \swebench{} follows a three-stage pipeline to ensure task quality: (1) scraping pull requests from well-maintained repositories, (2) filtering for those that resolve a specific issue and modify corresponding tests, and (3) retaining only instances that pass a "fail-to-pass" execution cycle, confirming the tests' relevance to the issue. In each task, an agent is provided with an issue description and the entire repository as the codebase, and is challenged to autonomously generate a corrective code patch. A submission is deemed successful only if the patch successfully applies and passes all unit and integration tests, which are hidden from the agent during the resolution task.
To facilitate more reliable evaluation and reduce evaluation overhead~\cite{agentless}, \swebenchverified{} provides a subset of 500 human-validated tasks with well-scoped tests and unambiguous issue descriptions~\cite{OpenAI2024SWEBench}.

Recently, researchers have extended the repository-level issue-resolution tasks for SWE-bench in a multilingual manner. For instance, Multi-SWE-bench~\cite{mswe} introduces a large-scale, human-annotated dataset of 1,632 tasks across seven languages, including Java, JavaScript, and Rust. SWE-bench Multilingual~\cite{khandpur2025multilingual} offers a more compact set of 300 high-quality tasks across nine languages, designed for rapid evaluation while maintaining full compatibility with the original SWE-bench infrastructure. SWE-PolyBench~\cite{swepoly} focuses on languages like Java and TypeScript. These benchmarks collectively represent a significant step towards evaluating the generalization capabilities of LLM-based agents across diverse software ecosystems. However, real-world Rust issues remain sparse in these benchmarks, featuring a small corpus of tasks (43 in SWE-bench Multilingual and 239 in Multi-SWE-bench \modify{vs. 500 in \swebenchverified{}}). Hence, there is an urgent need to construct a large-scale Rust real-world SWE benchmark to facilitate a rigorous and systematic evaluation of agent capabilities within the Rust ecosystem.

\section{\rustbench{}}

\begin{figure}[htb]
    \centering
    \includegraphics[width=0.97\columnwidth]{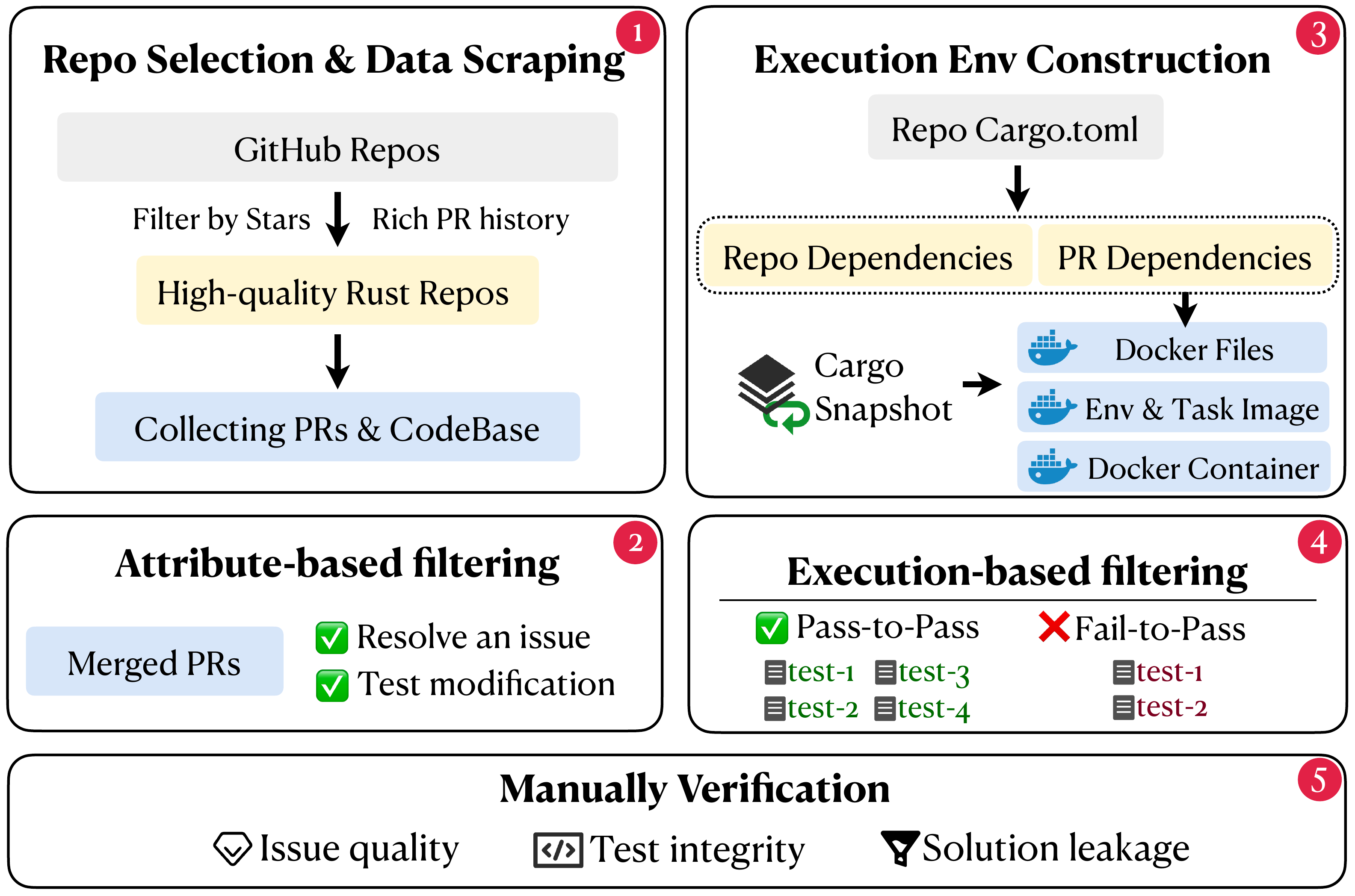}
    \caption{\rustbench{} Construction Process}
    \label{fig:construct-bench}
\end{figure}

\subsection{Benchmark Construction}
Figure~\ref{fig:construct-bench} shows a five-step construction process of \rustbench{}, delineated as follows. 
\subsubsection{Repository selection and data scraping}

We first select \repoCNT{} popular open-source Rust repositories that meet several criteria: they must have over \githubstar{} GitHub stars, be well-maintained, feature an extensive test suite, and possess a rich history of pull request (PR) discussions. Specifically, we scrape approximately \prCNT{} merged PRs from these repositories, along with the codebase state at each PR's corresponding base commit. 

\subsubsection{Attribute-based filtering}
We then apply an attribute-based filter to the scraped PRs, retaining only those that meet two criteria: being explicitly linked to a GitHub issue and introducing modifications to the project's test suite. Specifically, we conceptually divide the changes within each pull request's patch into two components: a test patch, which modifies the test suite, and a fix patch, which resolves the corresponding issue.

\subsubsection{Execution environment construction}
To ensure build reproducibility and mitigate dependency conflicts, 
we construct a snapshot-based execution environment for each task by configuring Cargo~\cite{rust-cargo-book} to use a crates.io index snapshot at the time of the original pull request.
Next, we analyze and classify dependencies as either repository-level or PR-specific to further minimize both storage overhead and the build time for each task's Docker image and container.

\subsubsection{Execution-based filtering}
Our validation process first \gr{identifies} the \texttt{fail} state of tests, i.e., we apply only the PR's test patch to the base commit and execute the test suite. Next, we apply the PR's fix patch and re-execute the test suite to identify the \texttt{pass} state. A task is validated as a genuine fail-to-pass instance only if it meets three conditions: (1) at least one test exhibits a clear transition from fail to pass, (2) no previously passing tests (pass-to-pass tests) regress, and (3) the complete patch (both test and fix) introduces no new build failures, runtime errors, or test regressions. Specifically, unlike Python, to rigorously define a \texttt{fail} in Rust's compiled environment, we perceive compilation errors as test failures. This ensures that any non-compiling state, whether after applying the test patch or the fix patch, is correctly categorized as a failure. Accordingly, we obtain a pool of \candidateTaskCNT{} candidate tasks.

\subsubsection{Manual verification}
We perform a rigorous manual verification of each candidate task instance, inspired by \swebenchverified{}~\cite{OpenAI2024SWEBench} and \agentless{}~\cite{agentless}. We assess each task's quality across three key dimensions: (1) issue quality, i.e., ensuring the problem description is sufficient and unambiguous; (2) fail-to-pass integrity, i.e., confirming the fail-to-pass behavior is deterministic and non-trivial; and (3) solution leakage, i.e., verifying that the issue does not contain explicit solutions. Our manual inspection details are provided on our GitHub page~\cite{githubrepo}. Notably, we label and classify each validated task by its primary type and challenge (as in Section~\ref{featurerustbench}). Eventually, our construction process yields the \rustbench{} dataset, which comprises 500 high-quality tasks from \selectedRepoCNT{} popular open-source Rust repositories, i.e., aligning with \modify{500 tasks in} the established \swebenchverified{}~\cite{OpenAI2024SWEBench} for Python, with each task representing a well-defined and realistic software engineering problem.

\newcommand{\reponame}[1]{\texttt{#1}}

\begin{table}[!t]
\caption{Characteristics of Repositories in \rustbench{}}
\centering
\label{tab:repo_summary_final}
\resizebox{0.98\columnwidth}{!}{%
\setlength\tabcolsep{8pt}
\begin{tabular}{@{}lccr@{}}
\toprule
\textbf{Repository} & \textbf{\#GitHub Stars} & \textbf{Functionality} & \textbf{\#Instance} \\
\midrule
\reponame{ripgrep} & 53.6k & Command-line search tool & 7 \\
\reponame{ast-grep} & 9.2k & CLI for code structural search & 6 \\
\reponame{async-graphql} & 3.5k & GraphQL server-side library & 14 \\
\reponame{cargo-dist} & 1.7k & Application packaging for Rust & 10 \\
\reponame{aya} & 3.7k & eBPF library for Rust & 6 \\
\reponame{bevy} & 40.5k & Data-driven game engine & 46 \\
\reponame{bincode} & 3.1k & Binary encoder/decoder & 8 \\
\reponame{biome} & 19.9k & Web project toolchain/linter & 53 \\
\reponame{boa} & 5.7k & Embeddable JavaScript engine & 49 \\
\reponame{cargo-generate} & 2.2k & Project template developer tool & 12 \\
\reponame{chrono} & 3.6k & Date and time library & 17 \\
\reponame{clap} & 15.3k & Command Line Argument Parser & 56 \\
\reponame{async-trait} & 2k & Async functions in traits & 9 \\
\reponame{autocxx} & 2.4k & Safe C++ from Rust interop & 5 \\
\reponame{bandwhich} & 10.6k & CLI network utilization display & 2 \\
\reponame{cargo-edit} & 3.2k & Manage cargo dependencies & 11 \\
\reponame{cbindgen} & 2.7k & Generates C bindings from Rust & 14 \\
\reponame{nushell} & 35.7k & A new type of shell & 5 \\
\reponame{rayon} & 11.9k & Data parallelism library & 2 \\
\reponame{angle-grinder} & 3.6k & Command-line log slicing tool & 3 \\
\reponame{askama} & 3.6k & Jinja-like template engine & 25 \\
\reponame{cargo-fuzz} & 1.7k & Command-line fuzzer for Rust & 6 \\
\reponame{cc-rs} & 2.0k & Compile C/C++ in build scripts & 8 \\
\reponame{chalk} & 1.9k & Rust trait system library & 7 \\
\reponame{cargo-make} & 2.8k & Task runner and build tool & 2 \\
\reponame{serde} & 9.8k & Serialization/deserialization & 1 \\
\reponame{bat} & 53.3k & A ‘cat’ clone with highlighting & 10 \\
\reponame{fd} & 38.6k & User-friendly alternative to ‘find’ & 12 \\
\reponame{cairo} & 1.8k & Language for provable programs & 9 \\
\reponame{axum} & 22.2k & Web application framework & 19 \\
\reponame{bytes} & 2.1k & Utilities for working with bytes & 22 \\
\reponame{tokio} & 29.0k & Asynchronous runtime for Rust & 17 \\
\reponame{tracing} & 6k & Application-level tracing & 13 \\
\reponame{burn} & 11.5k & Flexible Deep Learning Framework & 14 \\
\bottomrule
\end{tabular}%
}
\end{table}

\subsection{Characteristics of \rustbench{}}
\label{featurerustbench}

\rustbench{} provides 500 high-quality, manually reviewed task instances from \selectedRepoCNT{} repositories, making it considerably larger and more diverse than other existing Rust SWE benchmarks like SWE-bench Multilingual~\cite{khandpur2025multilingual} (43 instances from 7 repositories) and Multi-SWE-bench~\cite{mswe} (239 instances from 10 repositories). To ensure broad functional coverage, \rustbench{} incorporates diverse project domains as shown in Table~\ref{tab:repo_summary_final}, including command-line utilities (e.g., \texttt{ripgrep}~\cite{repo:ripgrep}, \texttt{fd}~\cite{repo:fd}), development tools (e.g., \texttt{cargo-edit}~\cite{repo:cargo-edit}, \texttt{biome}~\cite{repo:biome}), concurrency and parallelism libraries (e.g., \texttt{tokio}~\cite{repo:tokio}, \texttt{rayon}~\cite{repo:rayon}), web frameworks (e.g., \texttt{axum}~\cite{repo:axum}), and specialized engines (e.g., \texttt{bevy}~\cite{repo:bevy}, \texttt{boa}~\cite{repo:boa}).

\begin{table}[!t]
\centering
\caption{Overall Statistics of the \rustbench{}}
\label{tab:dataset_stats}
\resizebox{0.98\columnwidth}{!}{%
\setlength\tabcolsep{16pt}
\begin{tabular}{@{}llrrr@{}}
\toprule
\textbf{Component} & \textbf{Metric} & \textbf{Mean} & \textbf{Median} & \textbf{Max} \\
\midrule
\multirow{1}{*}{\textbf{Issue Text}} & Length (Tokens) & 393.4 & 245.5 & 3,826 \\
\midrule
\multirow{2}{*}{\textbf{Codebase}} & \# Files & 993.6 & 335 & 15,036 \\
 & \# Lines& 128,126 &  66,069 & 753,715 \\
\midrule

\multirow{3}{*}{\textbf{Fix Patch}}

 & \# Files edited & 9.8 & 4 & 148\\

 & \# Hunks edited & 9.9 & 4.5 & 148 \\

 & \# Lines edited & 139.9 & 40.5 & 15,449 \\

\midrule

\multirow{2}{*}{\textbf{Tests}} & \# Fail to Pass & 96.5 & 5 & 1,733 \\
& \# Pass to Pass & 219.6 & 14 & 1,780 \\
\bottomrule
\end{tabular}%
}
\end{table}

Table~\ref{tab:dataset_stats} presents the overall statistics of \rustbench{}, highlighting its scale and task complexity. Specifically, the \rustbench{} demonstrates significant diversity in codebase size. On average, its projects contain 993.6 files and 128,126 lines of code, with the largest repository reaching over 15,000 files and 750,000 lines. This scale ensures that \rustbench{} effectively represents a wide spectrum of real-world software development scenarios. Moreover, a key characteristic of \rustbench{} is the high complexity of its tasks, requiring an average of 9.8 files, 9.9 hunks, and 139.9 lines of code to resolve an issue, whereas a fix in Python-based \swebenchverified{}~\cite{OpenAI2024SWEBench} requires on average 1.25 files, 2.46 hunks, and 14.32 lines. This contrast implies \rustbench{} potentially better reflects real-world software engineering complexities.

\begin{figure}[t]
    \centering
    \includegraphics[width=0.74\columnwidth]{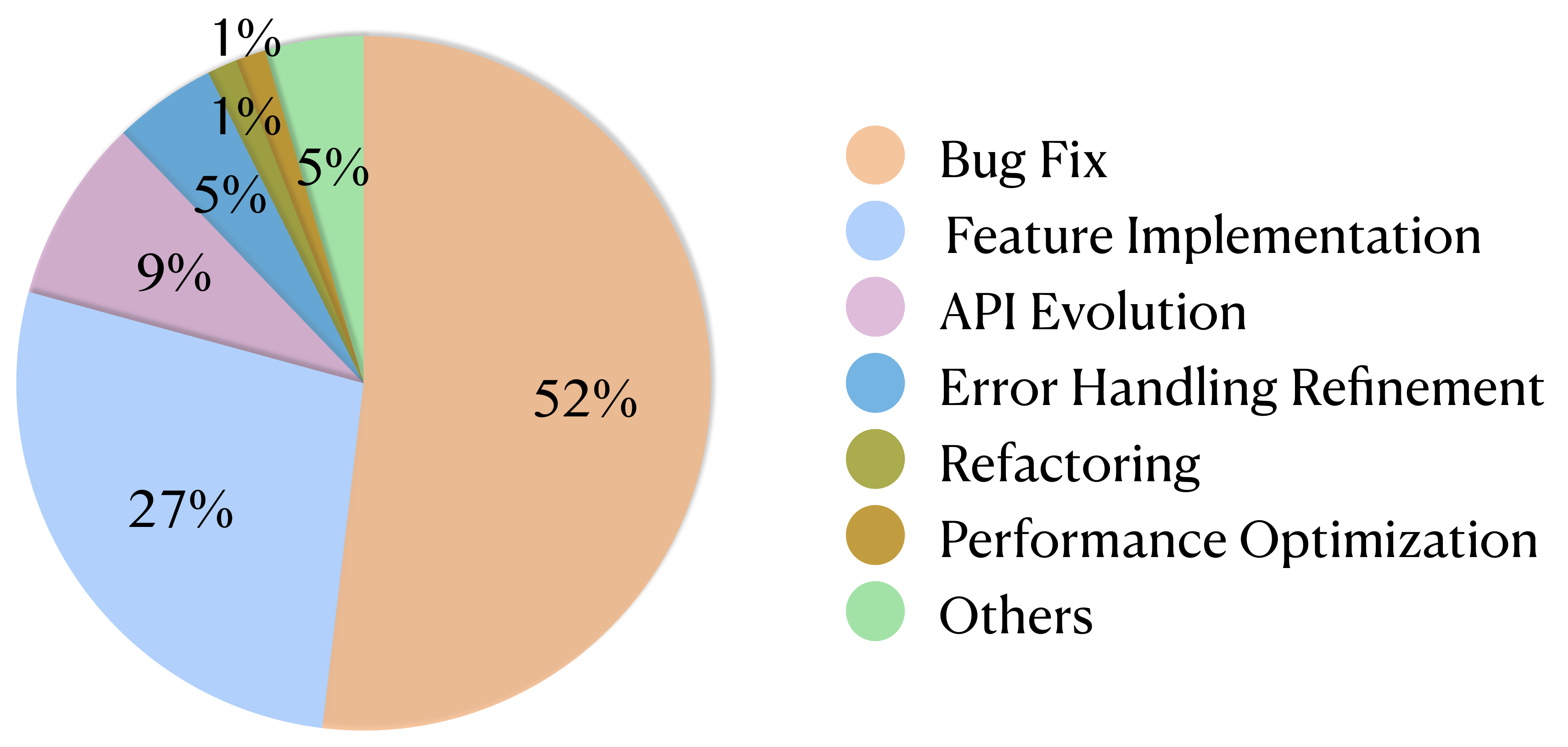}
    \caption{Distribution of Task Categories in \rustbench{}}
    \label{fig:rustbenchfeatures}
\end{figure}

To characterize the nature of the tasks, we manually classified all 500 instances. Figure~\ref{fig:rustbenchfeatures} shows their distribution where 52\% of the tasks are Bug Fix issues, followed by 27\% Feature Implementation issues, indicating that \rustbench{} mainly focuses on practical software maintenance and evolution workflows.
The high code churn detailed in Table~\ref{tab:dataset_stats} (averaging 139.9 lines edited \modify{vs. 14.3 in \swebenchverified{}}) suggests that complex code modifications are a common feature of routine maintenance in Rust. The benchmark also includes categories that are highly relevant to Rust's ecosystem, e.g., API Evolution (9\%) and Error Handling Refinement (5\%), reflecting the language's evolving APIs and emphasis on explicit error management~\cite{rustevo}.
Note that their original manual resolution takes on average 113 days from issue creation to PR merge and involves approximately 7.2 rounds of developer discussion.
\cmr{We further analyze the distribution of these resolution durations and find that 14.0\% of tasks are resolved within a day, 26.6\% within a week, 21.6\% within a month, and 21.8\% up to six months, while 16.0\% exceed six months. This indicates that \rustbench{} covers real-world scenarios, spanning from quick fixes to long-standing issues.}

\section{The Extensive Study}

\subsection{Study Setup}

\subsubsection{Study Subjects}

We evaluate four representative SWE \modify{agents} including \sweagent{}~\cite{sweagent}, \openhands{}+CodeAct v2.1~\cite{openhands}, \agentless{}~\cite{agentless}, and \autocode{} v2.0~\cite{specrover} on our \rustbench{} benchmark with their details as follows. 

\begin{itemize}[leftmargin=*]
    \item \sweagent{}: \sweagent{} designs an agent-computer interface which defines the possible actions taken by an agent to edit code, navigate the codebase, and execute tests.
    \item \openhands{}+CodeAct v2.1: \openhands{} utilizes the CodeAct~\cite{codeact} architecture, enabling it to perform tasks by executing code-based actions.
    \item \agentless{}: \agentless{} introduces a structured, multi-stage workflow (e.g., localize, repair, and validate) that deliberately constrains the LLM's autonomy to a predefined sequence.
    \item \autocode{} v2.0: \autocode{} v2.0 employs an iterative process of context retrieval and specification inference to progressively guide the generation of code patches.

\end{itemize}

\subsubsection{LLMs} 
We adopt four advanced LLMs including \claudesonnet{} (claude-3-7-sonnet-20250219)~\cite{anthropic2024claude3_7}, \gpto{} (gpt-4o-2024-11-20)~\cite{openai2024gpt4o}, \omini{} (o4-mini-2025-04-16)~\cite{openai2025o3o4mini}, and \qwen{} (qwen3-235b-a22b)~\cite{Qwen_Pricing}. We obtain the open-source model \qwen{} from Hugging Face~\cite{hugging-face-fc} and access \claudesonnet{}, \gpto{}, \omini{} through the APIs provided by Anthropic~\cite{anthropic_claude_docs_2025} and OpenAI~\cite{openai-api-fc}. Inference for the open-source model \qwen{} is conducted on servers with 128-core 2.6GHz AMD EPYC™ ROME 7H12 CPU, 512 GiB RAM, and eight NVIDIA A100 80GB GPUs, running Ubuntu 20.04.6 LTS, utilizing vllm~\cite{kwon2023efficient} inference framework.

\subsubsection{Evaluation Metrics}
\label{sec:study-eva-metrics}
Following prior works~\cite{swebench, mswe, swelancer, hu2025assessingadvancingbenchmarksevaluating}, our major evaluation metric is the \textbf{Resolved Rate (\%)}, reported as \textbf{Pass@1} efficacy. An issue is considered resolved if a single generated patch successfully applies to the codebase and passes all developer-written acceptance tests. Crucially, these acceptance tests are held out and not used by the agent during the patch generation process to ensure a fair evaluation. Moreover, we report the average API inference cost (\textbf{\$ Avg. Cost}) and token usage (\textbf{\# Avg. Token}).
Note that to calculate the cost of the open-source LLM \qwen{}, we adopt the pricing model from its official API webpage~\cite{Qwen_Pricing}. Furthermore, to specifically evaluate the issue reproduction stage adopted by all our four studied agents in RQ2, we introduce an additional metric: \textbf{Reproduction Success Rate (\%)}, which measures the ratio of tasks where the agent-generated test successfully replicates the behavior described in the original issue. 


\subsection{Implementations}

\subsubsection{Agent Adaptation for Rust}
Following the prior work~\cite{mswe}, we adapt the prompts of our four studied agents for the Rust language. Moreover, agents like \autocode{} and \agentless{} include an issue reproduction stage. However, reproducing issues in Rust requires using shell commands to manage the environment and dependencies of Cargo projects. Therefore, we develop custom shell tools to enable this stage for these agents on \rustbench{}. Furthermore, we adapt \autocode{}'s program structure-aware APIs from Python to Rust, enabling the agent to gather relevant code context. \cmr{For other configurations, including decoding parameters (e.g., temperature and top-$p$), we adhere to the original setups.} The detailed implementations are shown in our GitHub page~\cite{githubrepo}.


\subsubsection{Evaluation of the Agent's Reproduction Stage}
\label{reproduce-impl-setup}
To assess the agents' capabilities in reproducing issues, we extract the necessary commands and files from their issue-resolving trajectories and re-execute them within a sandboxed Docker container to capture the execution results. Specifically, we first run the reproduction test to verify its syntactic validity~\cite{swt-bench,aegis}. However, due to the complexity of Rust issues~\cite{reportquality}, we cannot analyze the reproduction results directly via assertions as in prior work~\cite{swt-bench,aegis}. Therefore, for tests meeting this criterion, we then assess the reproduction results via manual analysis of the test execution output (including error messages and exit codes) to determine if the observed behavior corresponds to the problem described in the original issue. 

{
\definecolor{tableshallowblue}{RGB}{229, 238, 251}
\definecolor{tablewhite}{RGB}{255,255,255}

\begin{table*}[!h]
    \centering
    \caption{Evaluation Results of Different Models and Agents in \rustbench{}}
    \setlength\tabcolsep{19pt}
    \label{tab:rq1-result}
    \resizebox{0.98\textwidth}{!}{%
    \begin{tabular}{l|l|rccccr}
    
    \toprule
    
    \multirow{2}{*}{\textbf{Agent}}
    & \multirow{2}{*}{\textbf{LLM}}
    & \multirow{2}{*}{\textbf{\% Resolved}}
    & \multicolumn{3}{c}{\textbf{\#Edited}}
    & \textbf{Avg.} & \multicolumn{1}{r}{\textbf{Avg.}} \\[1pt]
    \cline{4-6}
    & & & \rule{0px}{1.1em} \textbf{Line} & \textbf{Hunk} & \textbf{File} & \textbf{\$ Cost} & \multicolumn{1}{r}{\textbf{\# Token}} \\

    \midrule
    
    \rowcolor{tableshallowblue}
    \cellcolor{tablewhite}
    & \claudesonnet{}   & 106(21.20\%)              & 146.03    & 5.59     & 3.84 & 3.81      & 1,236,942 \\
    & \gpto{}           & 33\phantom{0}(6.60\%)     & 66.59     & 3.19     & 2.50 & 1.85      & 729,393  \\
    \rowcolor{tableshallowblue}
    \cellcolor{tablewhite}
    & \omini{}          & 34\phantom{0}(6.80\%)     & 20.25     & 2.32     & 1.42 & 1.23      & 1,106,811 \\
    \multirow{-4}{*}{\textbf{\openhands{}}}
    & \qwen{}           & 25\phantom{0}(5.00\%)     & 81.05     & 3.00     & 2.30 & 0.35      & 492,141  \\
    
    \midrule
    
    \rowcolor{tableshallowblue}
    \cellcolor{tablewhite}
    & \claudesonnet{}   & 75(15.00\%)               & 190.49    & 6.71     & 4.02 & 3.48      & 1,131,753 \\
    & \gpto{}           & 9\phantom{0}(1.80\%)      & 172.84    & 5.28     & 3.45 & 1.92      & 755,343  \\
    \rowcolor{tableshallowblue}
    \cellcolor{tablewhite}
    & \omini{}          & 41\phantom{0}(8.20\%)     & 77.13     & 3.81     & 2.32 & 2.13      & 1,708,013 \\
    \multirow{-4}{*}{\textbf{\sweagent{}}}
    & \qwen{}           & 9\phantom{0}(1.80\%)      & 83.47     & 2.33     & 1.72 & 0.33      & 417,091  \\
    
    \midrule
    \rowcolor{tableshallowblue}
    \cellcolor{tablewhite}
    & \claudesonnet{}   & 36\phantom{0}(7.20\%)     & 13.03     & 1.37     & 1.06 & 1.47      & 439,850  \\
    & \gpto{}           & 27\phantom{0}(5.40\%)     & 16.45     & 1.93     & 1.24 & 0.89      & 331,373  \\
    \rowcolor{tableshallowblue}
    \cellcolor{tablewhite}
    & \omini{}          & 39\phantom{0}(7.80\%)     & 13.10     & 1.43     & 1.06 & 0.61      & 441,949  \\
    \multirow{-4}{*}{\textbf{\agentless{}}}
    & \qwen{}           & 16\phantom{0}(3.20\%)     & 13.41     & 1.30     & 1.07 & 0.35      & 460,175  \\
    
    \midrule
    \rowcolor{tableshallowblue}
    \cellcolor{tablewhite}
    & \claudesonnet{}   & 46\phantom{0}(9.20\%)     & 22.01     & 1.45     & 1.13 & 1.18      & 352,467  \\
    & \gpto{}           & 21\phantom{0}(4.20\%)     & 15.01     & 1.53     & 1.24 & 1.09      & 369,833  \\
    \rowcolor{tableshallowblue}
    \cellcolor{tablewhite}
    & \omini{}          & 34\phantom{0}(6.80\%)     & 17.38     & 1.51     & 1.09 & 0.46      & 343,247  \\
    \multirow{-4}{*}{\textbf\autocode{}}
    & \qwen{}           & 24\phantom{0}(4.80\%)     & 12.09     & 1.37     & 1.14 & 0.32      & 359,351  \\
    
    \bottomrule
    \end{tabular}
    }
\end{table*}

}

\subsection{Research Questions}

We investigate the following research questions to study the effectiveness of the agentic approaches and the factors that impact their effectiveness on \rustbench{}.

\begin{itemize}[leftmargin=*]
    \item \parabf{RQ1:} \textit{How do different agents perform on \rustbench{}?} 
    For this RQ, we benchmark the selected agents-model configurations on \rustbench{} and report their overall effectiveness.

    \item \parabf{RQ2:} \textit{What are the behavioral characteristics of agents resolving real-world Rust issues?} 
    For this RQ, we perform a detailed behavioral analysis of the agents, which involves examining the scope of code edits, the types of compilation errors they produce, and the tests they reproduce.

\end{itemize}

\subsection{Result Analysis}

\subsubsection{RQ1: performance of agents and models on \rustbench{}}

Table~\ref{tab:rq1-result} presents the resolution rates for each agent-model configuration on \rustbench{}. 
We find that agents built upon the ReAct paradigm~\cite{react}, such as \openhands{} and \sweagent{}, achieve the best performance on this benchmark. 
Specifically, the top-performing agent-model configuration, \openhands{} paired with \claudesonnet{}, resolves 106 (21.2\%) tasks. When resolved manually, these tasks typically require an average of 126 days from issue opening to PR merge and 5.5 rounds of discussion. Notably, this result largely \gr{outperforms} not only the runner-up ReAct-based agent \sweagent{} (15.0\% tasks), but also the runner-up LLM with its own framework \omini{} (6.8\%). These results underscore the decisive and compounding impact of both the ReAct-style agentic architecture and the choice of the underlying language model.

\mybox{Finding 1: The adoption of a ReAct-style architecture achieves the best performance on \rustbench{}.}



Moreover, we observe a significant cost disparity among the agent-model configurations. The top-performing agents, \openhands{} and \sweagent{}, are also the most expensive, e.g., costing \$3.81 and \$3.48 on average  when paired with \claudesonnet{}, respectively. This higher cost is potentially caused by their ReAct-based~\cite{react}, iterative problem-solving frameworks, which lead to a greater number of interaction rounds and consequently higher token consumption compared to the more structured, single-pass workflows of \agentless{} and \autocode{}.



\begin{figure}[htb]
    \centering
    \includegraphics[width=0.74\columnwidth]{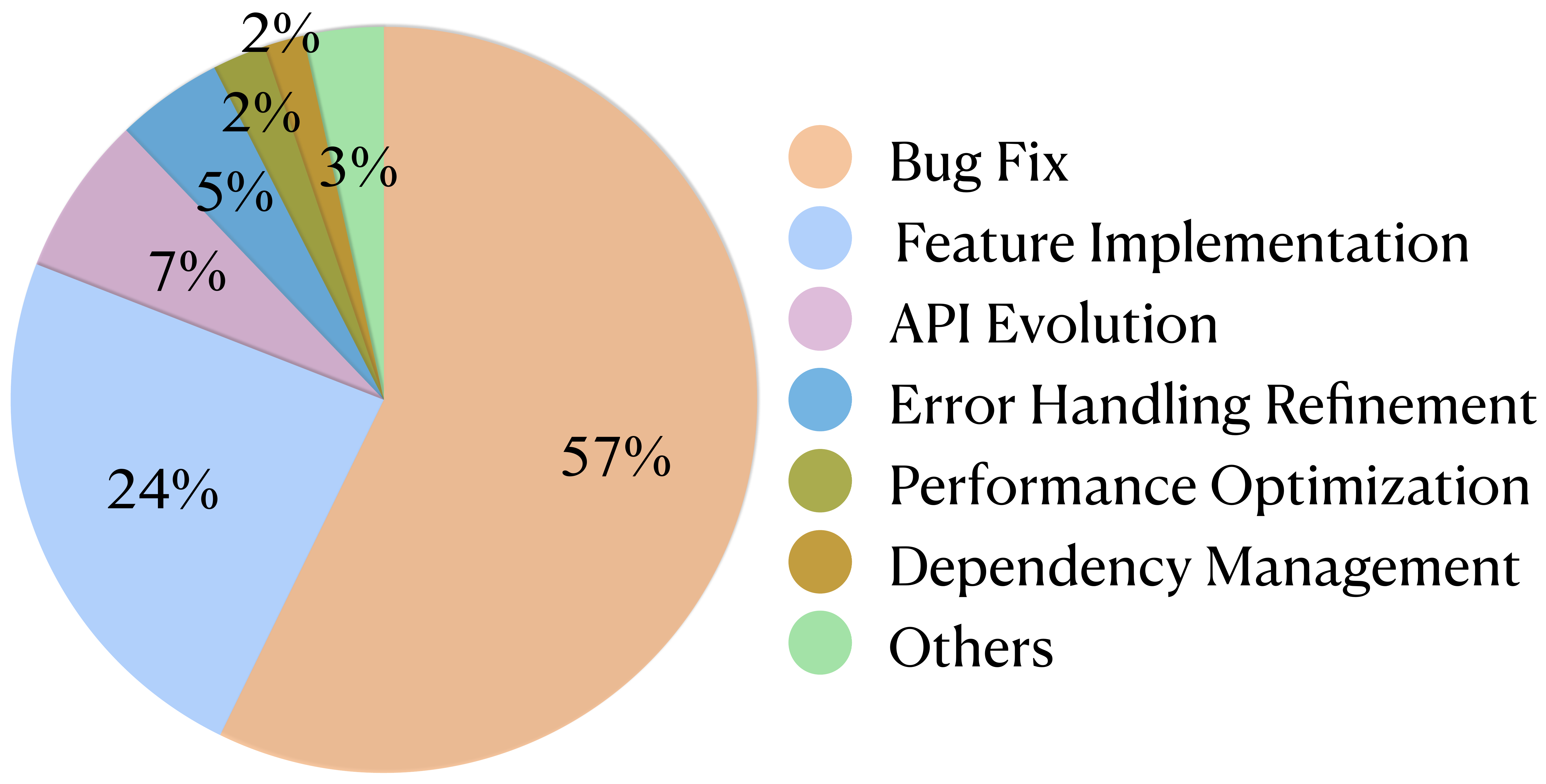}
    \caption{Distribution of All Resolved Tasks by Category.}
    \label{fig:task-distribution-by-category}
\end{figure}
Figure~\ref{fig:task-distribution-by-category} shows the distribution of the resolved task types by all studied agents which closely aligns with the overall benchmark's composition in Figure~\ref{fig:rustbenchfeatures}, e.g., bug fixes (57\%) and feature implementations (24\%) dominate the resolved tasks. This alignment demonstrates that the agents are effective in tackling common challenges in the Rust ecosystem. 

\subsubsection{RQ2: behavioral analysis of agents on \rustbench{}}


\modify{Table~\ref{tab:rq1-result} presents the} edit scopes of patches generated by different agents. We find that agents like \sweagent{} and \openhands{} cause substantially larger modifications, editing on average over 2 files and 70 lines of code per task, while \agentless{} and \autocode{} typically edit 1.2 files on average and fewer than 20 lines. 

\begin{figure}[htb]
    \centering
    \includegraphics[width=0.85\columnwidth]{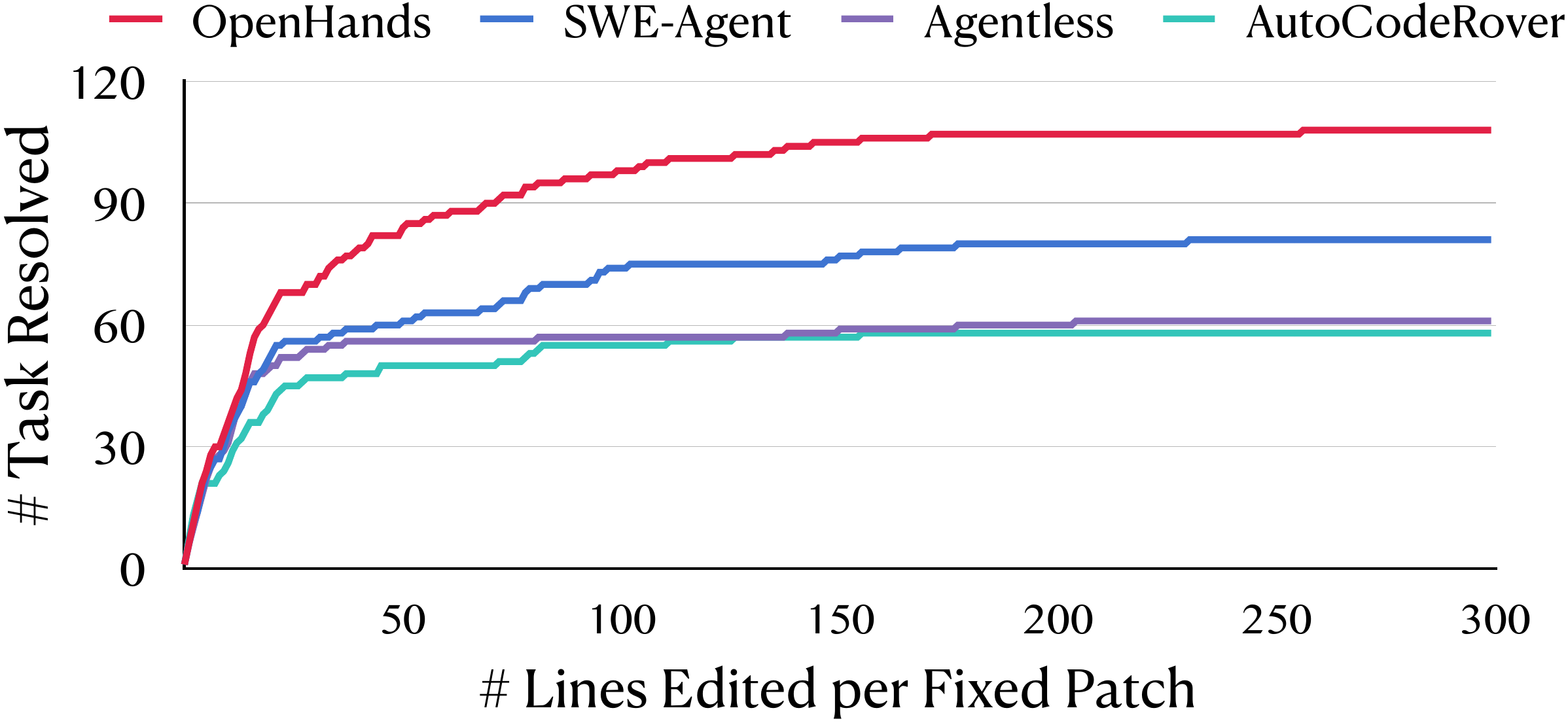}
    \caption{Cumulative Resolved Tasks by \cmr{Edited Lines per Fixed Patch}, Aggregated per Agent Across All LLMs.}
    \label{fig:edit-trend}
\end{figure}

We further analyze the capabilities of different agents on tasks requiring varying edit scopes, with results aggregated across all models for each agent, as shown in Figure~\ref{fig:edit-trend}. The result reveals that all four agents perform similarly on tasks solvable with small patches, e.g., \openhands{} resolves 48 tasks and \agentless{} resolves a comparable 43 within a 15-line edit scope. In contrast, a clear performance gap appears as the required patch size increases. Within the 150-line scope, \openhands{} resolves 105 tasks, while \agentless{} resolves only 57. Such results indicate that powerful agents like \openhands{} and \sweagent{} tend to explore a wider solution space by making more extensive changes. 

\mybox{Finding 2: While all agents are similarly effective on tasks requiring small patches, top-performing agents are more powerful to generate large, complex patches for more challenging tasks.}

{
    \renewcommand{\tabularxcolumn}[1]{>{\raggedright\arraybackslash}m{#1}} 

    \newcommand{\bhref}[3][blue]{\href{#2}{\color{#1}{#3}}}%
    
    \begin{table}[ht]
        \centering
        \caption{Statistics of Common Rust Compilation Errors Encountered During Agent Issue Resolution}
        \setlength\tabcolsep{8pt}
        \label{tab:c11_e5}
        \resizebox{0.98\columnwidth}{!}{%
        \begin{tabularx}{1.35\columnwidth}{c|X|r}
    
        \toprule
    
        \textbf{Error Code} & \textbf{Error Description} & \textbf{Rate (\%)} \\
    
    \midrule
    \midrule

    \bhref{https://doc.rust-lang.org/error_codes/E0599.html}{\texttt{E0599}} & A method is used on a type which doesn't implement it & 18.06\% \\
    \midrule
    \bhref{https://doc.rust-lang.org/error_codes/E0433.html}{\texttt{E0433}} & An undeclared crate, module, or type was used & 16.21\% \\
    \midrule
    \bhref{https://doc.rust-lang.org/error_codes/E0432.html}{\texttt{E0432}} & An import was unresolved & 12.08\% \\
    \midrule
    \bhref{https://doc.rust-lang.org/error_codes/E0425.html}{\texttt{E0425}} & An unresolved name was used & 8.54\% \\
    \midrule
    \bhref{https://doc.rust-lang.org/error_codes/E0308.html}{\texttt{E0308}} & Expected type did not match the received type & 6.81\% \\
    \midrule
    \bhref{https://doc.rust-lang.org/error_codes/E0277.html}{\texttt{E0277}} & Type does not implement expected trait & 6.50\% \\
    \midrule
    \bhref{https://doc.rust-lang.org/error_codes/E0412.html}{\texttt{E0412}} & A used type name is not in scope & 5.69\% \\
    \midrule
    \bhref{https://doc.rust-lang.org/error_codes/E0753.html}{\texttt{E0753}} & An inner doc comment was used in an invalid context & 3.07\% \\
    \midrule
    \bhref{https://doc.rust-lang.org/error_codes/E0282.html}{\texttt{E0282}} & The compiler could not infer a type and asked for a type annotation & 1.75\% \\
    \midrule
    \bhref{https://doc.rust-lang.org/error_codes/E0609.html}{\texttt{E0609}} & Attempted to access a nonexistent field in a struct & 1.48\% \\
    \midrule
    \bhref{https://doc.rust-lang.org/error_codes/E0061.html}{\texttt{E0061}} & An invalid number of arguments was passed when calling a function & 1.23\% \\
    \midrule
    \bhref{https://doc.rust-lang.org/error_codes/E0405.html}{\texttt{E0405}} & The code refers to a trait that is not in scope & 1.21\% \\
    \midrule
    \bhref{https://doc.rust-lang.org/error_codes/E0407.html}{\texttt{E0407}} & A definition of a method not in the implemented trait was given in a trait implementation & 1.18\% \\
    
    \bottomrule
        
        \bottomrule
        
        \end{tabularx}
        }
    \end{table}
}

\begin{figure}[htb]
    \centering
    \includegraphics[width=0.97\columnwidth]{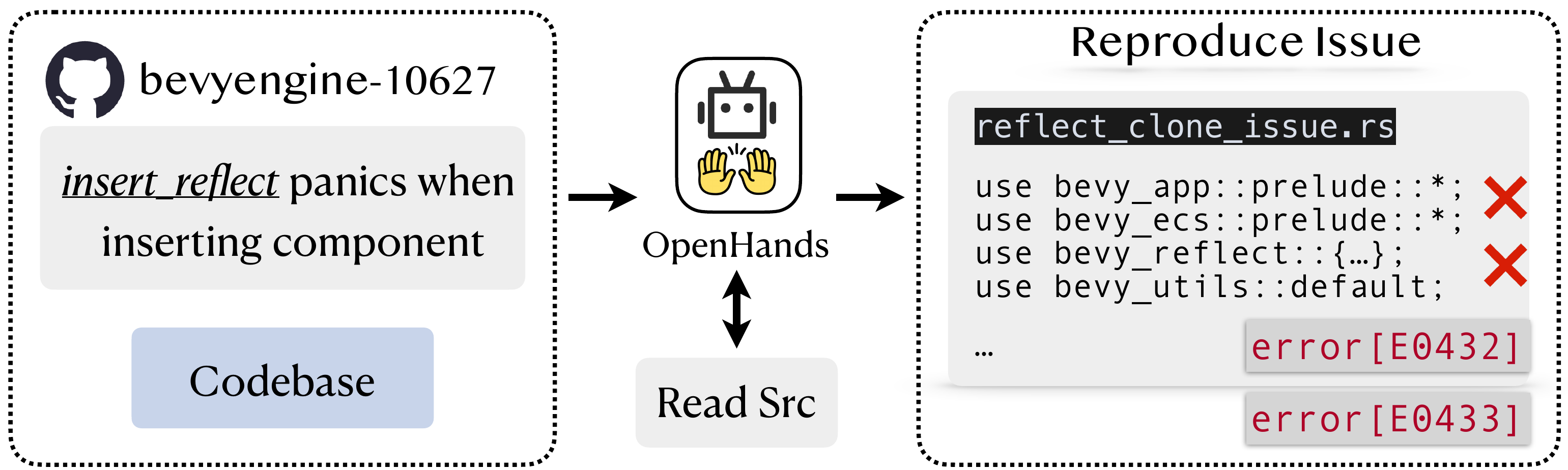}
    \caption{Compilation Error in \texttt{bevyengine-10627} Task}
    \label{fig:bevy}
\end{figure}

We find agents frequently encounter compilation errors when resolving real-world Rust issues, e.g., \openhands{} encounters an average of three compilation errors per task. We then examine the overall distribution of common compilation errors produced during their resolution process, detailed in Table~\ref{tab:c11_e5}. We find that the compilation errors originate from two major challenges. One is the failure to comprehend repository-wide code organization (43.7\%), leading to errors in naming, scoping, and path resolution (i.e., \texttt{E0433}, \texttt{E0432}, \texttt{E0425}, \texttt{E0412}, and  \texttt{E0405}). This suggests that agents struggle to correctly model the project's structural context for linking different code modules, which is a prominent challenge during the issue reproduction stage. For instance, in the case of \texttt{bevyengine-10627}~\cite{traj_openhands_claude_evyengine_bevy_10627} shown in Figure~\ref{fig:bevy}, \openhands{} with \claudesonnet{} attempts to construct a reproduction test to replicate a panic~\cite{rust_book_2024} where cloning a reflected trait object (\texttt{Box<dyn Reflect>}) via the \texttt{clone\_value()} method erases its type information, causing a subsequent call to \texttt{insert\_reflect} to fail. However, compilation errors arise from unresolved imports and undeclared types, causing a failure to incorporate the necessary \texttt{bevy crate preludes}~\cite{repo:bevy} and dependencies into the test's scope.

Another major challenge for compilation errors (32.6\%) arises from Rust's complex type and trait system. While agents often generate syntactically plausible code, they fail to satisfy the strict semantic contracts of Rust (i.e., \texttt{E0599}, \texttt{E0308}, \texttt{E0277}, and \texttt{E0407}). For example, in the case of \texttt{askama-374}~\cite{traj_sweagent_o4mini_rinjars_askama_374}, \sweagent{}'s syntactically valid attempt to call a method on an object consistently triggered an \texttt{E0599} compilation error  because the object's type failed to satisfy the necessary trait bounds, indicating the agent's difficulty to comply with Rust's strict semantics. 

\mybox{Finding 3: Agent-generated compilation errors primarily stem from two key deficiencies: a failure to model repository-wide code structure and an inability to comply with Rust's strict type and trait semantics.}

\begin{figure}[htb]
    \centering
    \includegraphics[width=0.94\columnwidth]{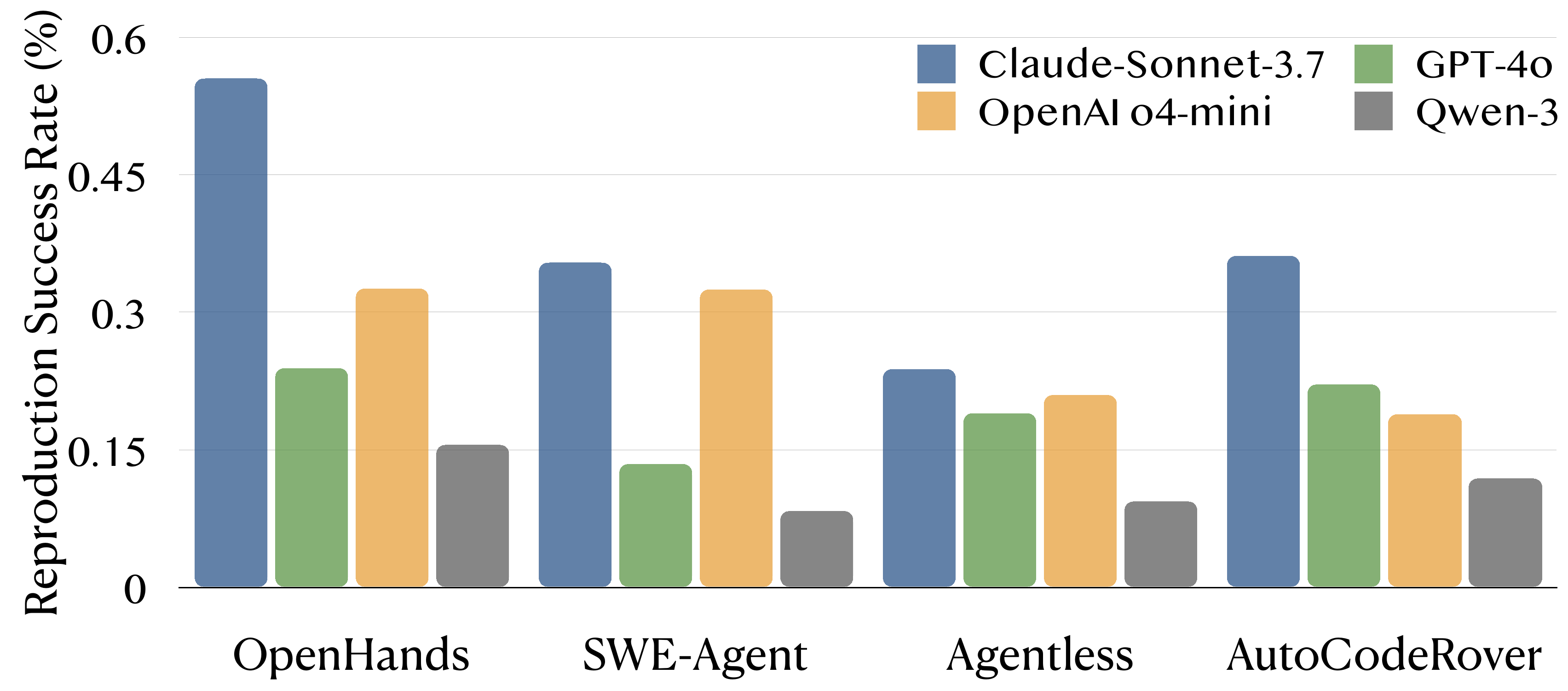}
    \caption{Reproduction Success Rate by Agents and Models} 
    \label{fig:reproduce-stage}
\end{figure}


Motivated by Finding 3, we find that these compilation errors largely occur in the issue reproduction stage, e.g., 35\% of them occur during the reproduction stage for the top-performing \openhands{}-\claudesonnet{} configuration, indicating that it is challenging for agents to reproduce issues.
Figure~\ref{fig:reproduce-stage} presents our evaluation of the issue reproduction stage, which involves re-executing tasks from \rustbench{} and analyzing their outputs. 
\openhands{} paired with \claudesonnet{} achieves the highest reproduction success rate at 55.5\%, i.e., the remaining 44.5\% of tasks fail at the reproduction stage, i.e., a critical prerequisite for validating patches and obtaining runtime information, and thus could not be successfully resolved even under the top-performing agent-model configuration.
Specifically, we find that for each successfully reproduced issue, \openhands{} with \claudesonnet{} re-runs the reproduction test on average 4.7 times for patch validation and dynamic information gathering. 
Moreover, we also disable the reproduction stage in \openhands{} paired with \claudesonnet{} and find that this causes a 42\% drop in its resolution rate, indicating that issue reproduction is a prominent bottleneck for Rust issue resolution. \cmr{To better understand these reproduction failures, we manually inspect the failed instances and find that most cases stem from dependency issues (52.5\%; misconfigured \texttt{Cargo.toml} or workspace paths), followed by irrelevant tests (35.5\%; tests that do not exercise the reported behavior) and non-compiling tests (12.0\%; generated tests that fail to build).}

\mybox{Finding 4: Issue reproduction is critical for Rust issue resolution.}

\section{Approach}
Inspired by Finding 4, we introduce \rustagent{}, a novel agentic approach to enhance Rust issue resolution through automated test environment setup coupled with a unique dynamic tracing strategy. 

\subsection{\rustagent{} Framework}

\begin{figure*}[!htb]
    \centering
    \includegraphics[width=\textwidth{}]{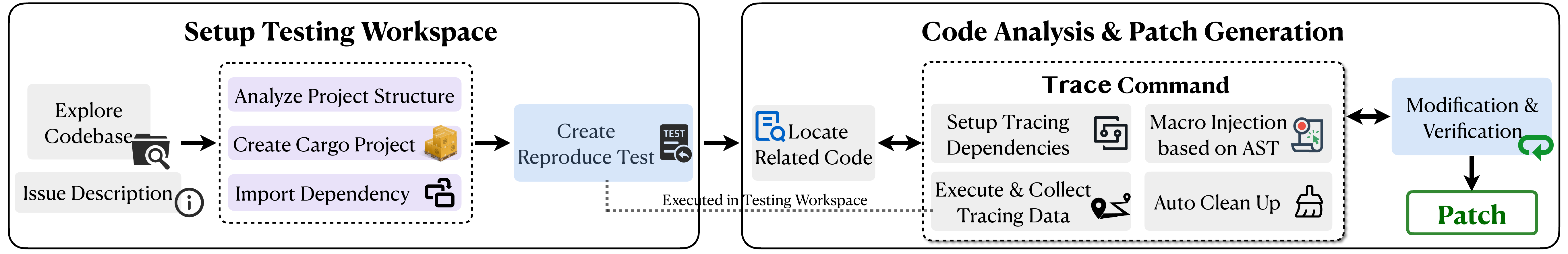}
    \caption{The \rustagent{} Framework}
    \label{fig:approach-framework}
\end{figure*}

Figure~\ref{fig:approach-framework} illustrates the framework of \rustagent{} comprising two stages. 
In the first stage \textit{Setup Testing Workspace}, the agent's primary objective is to create a reliable, isolated environment for issue reproduction.
It begins by parsing the input issue description and performing an initial exploration of the target codebase to understand its module structure and dependencies\cmr{, adhering to the established practices of ReAct-style agents~\cite{sweagent, openhands}}.
Subsequently, the agent automatically executes three key steps: (1) it analyzes the project's \texttt{Cargo.toml} to identify its structure and dependencies; (2) it initializes a new, clean Cargo project in a separate workspace; and (3) it imports the original, target project as a local path dependency into this new workspace. This ensures that any test will run against the exact code under investigation. Finally, within this isolated workspace, the agent generates a test case to specifically trigger and reproduce the failure described in the issue.

Once the issue is successfully reproduced, the agent launches the second stage, \textit{Code Analysis \& Patch Generation}, where it leverages the controlled environment for dynamic analysis. After using the test outputs to locate potential issue-related code regions, it activates its core cross-project dynamic tracing capability \modify{via the \texttt{Trace} command, a unified agent-computer interface that encapsulates the entire analysis workflow into a single, all-in-one function}. This process is highly automated, powered by Rust metaprogramming features~\cite{rl,rl_macro1,rl_ast1}: \rustagent{} first injects the necessary tracing dependencies into the test environment. It then instruments targeted functions within the original codebase by injecting tracing macros via Abstract Syntax Tree (AST) modification. Critically, the test execution and data collection are launched from the testing workspace, as indicated by the dotted arrow in Figure~\ref{fig:approach-framework}. This decoupled strategy allows the agent to capture precise, runtime control-flow information without being entangled in the original project's complex build system or test suite. The collected trace data provides the potential path to the issue's origin, enabling the agent to propose a code modification. This patch is then iteratively verified against the reproduction test until the issue is resolved and the agent outputs the final patch. \modify{Notably, \rustagent{}'s design is hybrid: it prioritizes the dynamic analysis powered by the \texttt{Trace} command for reproducible issues, while utilizing the static analysis for all other cases,} i.e., the agent bypasses dynamic analysis and proceeds directly to the \modify{\textit{Code Analysis \& Patch Generation} stage}, relying solely on static analysis to locate and modify the code without using the \texttt{Trace} command.

\subsection{\texttt{Trace} Command}

In this section, we introduce the \texttt{Trace} command for root cause analysis in complex Rust projects. It provides a programmatic interface to a cross-project tracing workflow, which automates code instrumentation, test execution, and runtime data collection. Specifically, the complete workflow of the \texttt{Trace} command integrates three key components: a selective AST-based instrumentor, a robust tracing runtime with intelligent data serialization, and a unified agent-computer interface for seamless agent integration.

\paragraph{Selective AST-based Instrumentation}
To capture execution flow and ensure syntactic integrity, \rustagent{}'s \texttt{Trace} command employs a selective instrumentor that directly modifies the source code's Abstract Syntax Tree (AST). It injects tracing macros at specified target functions. 
Specifically, the process is decoupled from the build system, enabling instrumentation of the original project code while running tests in a separate, isolated workspace. \modify{To ensure robustness and manage complexity, our current implementation of the \texttt{Trace} command focuses on instrumenting regular functions, while automatically excluding more complex structures like closures, other procedural macros, and test functions to prevent potential tracing conflicts and irrelevant overhead.}

\paragraph{Robust Tracing Runtime}
Our tracing runtime reconstructs the dynamic call graph by capturing function input and output. However, this task is challenging due to Rust's type system, in which the non-serializability~\cite{serde_serialize} of many types would otherwise cause compilation failures. To handle this, we adopt a hybrid tracing strategy that captures serializable types as JSON while recording descriptive placeholders for others, thereby preserving crucial type information without compromising build integrity. The resulting trace, encompassing the call hierarchy and captured data, is structured into a single JSON object to facilitate programmatic analysis.

\paragraph{Unified Agent-Computer Interface}
The \texttt{Trace} command encapsulates our entire tracing workflow into a unified agent-computer interface, inspired by \sweagent{}~\cite{sweagent}. The agent invokes it with parameters defining the target instrumentation functions, execution command, the testing workspace path, and the target codebase path. To guarantee a non-destructive operation, the \texttt{Trace} command automatically backs up the project state before instrumentation and restores it upon completion\modify{, usually within a second}. \modify{This interface abstracts the underlying complexity, i.e., \modify{automating dependency setup, AST-based macro injection, test execution, data collection, and final cleanup (Figure~\ref{fig:approach-framework}),} enabling the agent to focus on high-level resolution strategies instead of implementation details.}

\subsection{Evaluation}

\subsubsection{Evaluation Metrics and Baselines}
To evaluate the performance of \rustagent{}, we adopt metrics used in our extensive study (Section~\ref{sec:study-eva-metrics}): Resolve Rate (\%), \$ Avg. Cost, and Reproduction Success Rate (\%). Due to budget constraints, we follow the \sweagent{} setup~\cite{sweagent} to set the budget per task to \$4.
We adopt the same four studied agents and LLMs in our previous study as our baselines. 

\subsubsection{Implementation}
\rustagent{} is implemented as a ReAct-style agent~\cite{react} and follows the setup of baseline agents~\cite{sweagent, specrover, openhands}, with its temperature set to 0 to ensure the model's output is more deterministic. 
\cmr{Notably, our \texttt{Trace}'s instrumentation mechanism is designed as a purely observational wrapper to strictly preserve original program semantics. To ensure identical trait resolution, the injected macros retain original function signatures, including all generics and \texttt{where} clauses. Furthermore, we adhere to Rust's borrow checker by exclusively utilizing temporary, strictly-scoped immutable references to capture parameter values. This ensures that variable lifetimes and drop timings remain unaffected, with all macro-generated variables isolated within private internal scopes.} \cmr{Specifically, \texttt{Trace}'s overhead is minimal: tokens generated for tracing account for only 1.65\% of total usage, the shared isolated workspace's initial build takes about 40 seconds, and each \texttt{Trace} invocation adds only \(\sim\)12 seconds of incremental compilation.} Due to page limits, implementation details like \rustagent{}'s prompt, basic functions (e.g., string replacer) and \texttt{Trace} are shown in our GitHub page~\cite{githubrepo}.

\subsubsection{Result Analysis}

\begin{table}[t]
    \centering
    \caption{Evaluation Result of \rustagent{}}
    \setlength\tabcolsep{0.8pt}
    \label{tab:approach-result}
    \resizebox{0.98\columnwidth}{!}{%
        \begin{tabular}{lcccc}
            \toprule
            \textbf{Agent} & \textbf{\claudesonnet{}} & \textbf{\gpto{}} & \textbf{\omini{}} & \textbf{\qwen{}} \\
            \midrule
            \openhands{}            & 106(21.2\%)           & 33(6.60\%)            & 34(6.80\%)            & 25(5.00\%) \\
            \sweagent{}             & 75(15.0\%)            & 9(1.80\%)             & 41(8.20\%)            & 9(1.80\%) \\
            \agentless{}            & 36(7.2\%)             & 27(5.40\%)            & 39(7.80\%)            & 16(3.20\%) \\
            \autocode{}             & 46(9.20\%)            & 21(4.20\%)            & 34(6.80\%)            & 24(4.80\%) \\
            \textbf{\rustagent{}}   & \textbf{143(28.6\%)}  & \textbf{42(8.4\%)}    & \textbf{82(16.4\%)}   & \textbf{33(6.60\%)}          \\
            \bottomrule
        \end{tabular}
    }
\end{table}

Table~\ref{tab:approach-result} presents the evaluation results where \rustagent{}, when paired with \claudesonnet{}, achieves the best performance on \rustbench{} by successfully resolving 143 out of 500 tasks (28.6\%), outperforming the best-performing agent in our study \openhands{} with \claudesonnet{} by 34.9\%. 
\modify{Notably, the \texttt{Trace} command is successfully leveraged in the resolution process for 71.3\% of these tasks.} \cmr{Moreover, we validate the statistical significance of this result using McNemar's test~\cite{mcnemar1947note}, a widely adopted statistical method for comparing paired binary outcomes in software engineering research. The test yields a $p\text{-value} < 0.001$ when comparing \rustagent{} with the runner-up \textsc{OpenHands}, indicating that the performance gain is statistically significant.}  \rustagent{} also consistently outperforms each baseline agent across all four tested LLMs.
This consistent superiority underscores the robustness and general effectiveness of our proposed framework in tackling real-world Rust issues. Notably, \rustagent{} with the cost-effective \omini{} achieves a 16.4\% resolution rate (82 tasks), surpassing not only all baselines using the same model but also the highly-regarded \sweagent{} paired with the much more powerful \claudesonnet{} (15.0\%). Such a result indicates that by effectively addressing the core challenges of issue reproduction and runtime analysis, our approach enables even less powerful models to achieve highly competitive results on complex real-world repository-level Rust tasks. At last, across all LLMs, \rustagent{} uniquely resolves \approachUQResolve{} tasks that no baselines could.

\begin{figure}[!htb]
    \centering
    \includegraphics[width=\columnwidth{}]{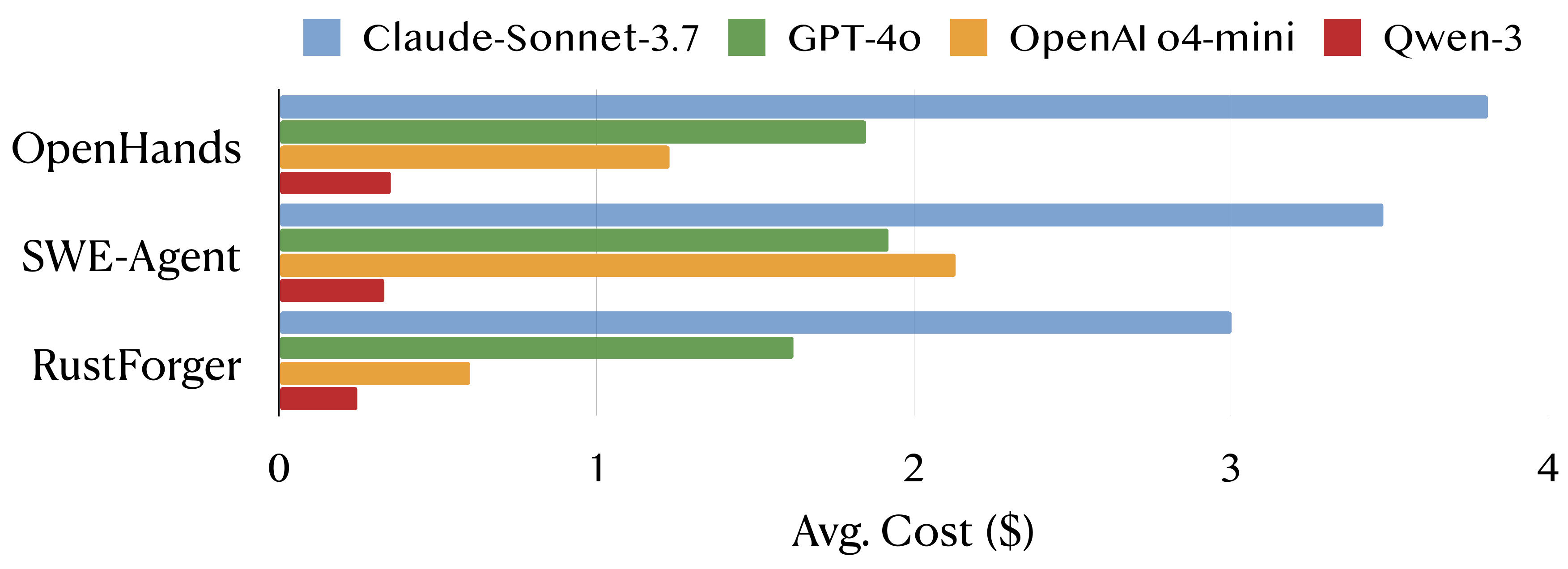}
    \caption{Comparison of Average Cost per Task}
    \label{fig:approach-cost}
\end{figure}

In addition to its superior resolution rates, \rustagent{} also demonstrates significant advantages in cost efficiency. As shown in Figure~\ref{fig:approach-cost}, \rustagent{} consistently incurs lower average API costs than the top-performing baselines, \openhands{} and \sweagent{}, across all LLMs. For instance, when paired with the most powerful model \claudesonnet{}, \rustagent{}'s average cost is \$3.0, which is 21.3\% and 13.8\% lower compared to \openhands{} (\$3.81) and \sweagent{} (\$3.48), respectively. This gain is even more significant with more economical models--with \omini{}, \rustagent{}'s cost (\$0.6) is less than half that of \openhands{} (\$1.23) and amounts to only 28.2\% of the cost of \sweagent{} (\$2.13). 

\cmr{Moreover, we examine how task complexity, approximated by the original manual resolution duration, correlates with agent performance. Our analysis shows that \rustagent{}’s success rate is 42.9\% on issues resolved within a day, 32.3\% within a week, 31.5\% within a month, 22.0\% within six months, and 15.0\% for issues open longer than six months, highlighting both \rustagent{}’s strong performance and the particular difficulty of long-standing issues.}

\begin{figure}[!htb]
    \centering
    \includegraphics[width=0.98\columnwidth{}]{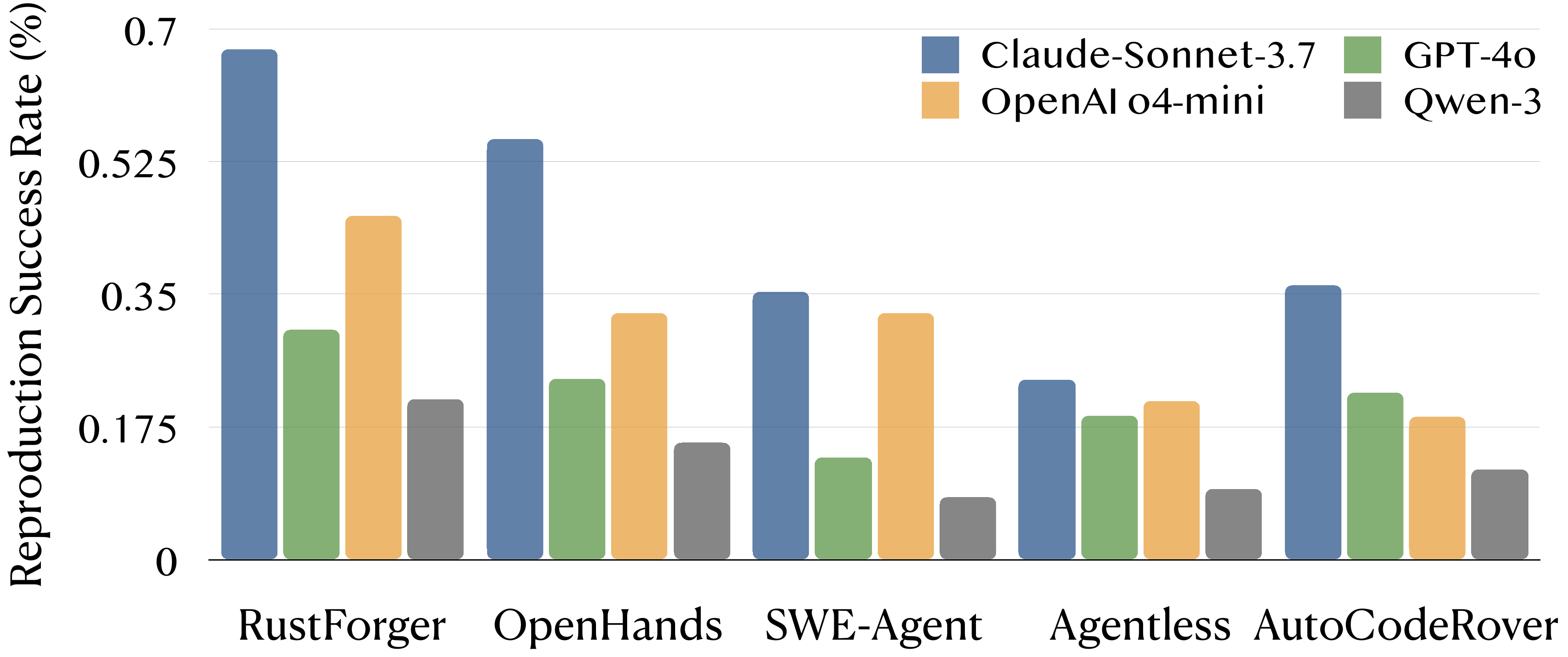}
    \caption{Comparison of Reproduction Success Rate}
    \label{fig:approach-reproduce}
\end{figure}

Figure~\ref{fig:approach-reproduce} shows that \rustagent{} consistently outperforms all baselines in terms of reproduction success rate across all LLMs, achieving a 67.3\% rate with \claudesonnet{} (vs. 55.5\% for \openhands{}) and a 45.4\% rate with \omini{} (vs. 32.6\% for \openhands{}).

\paragraph{Case Study}

\begin{figure}[!htb]
    \centering
    \includegraphics[width=0.95\columnwidth{}]{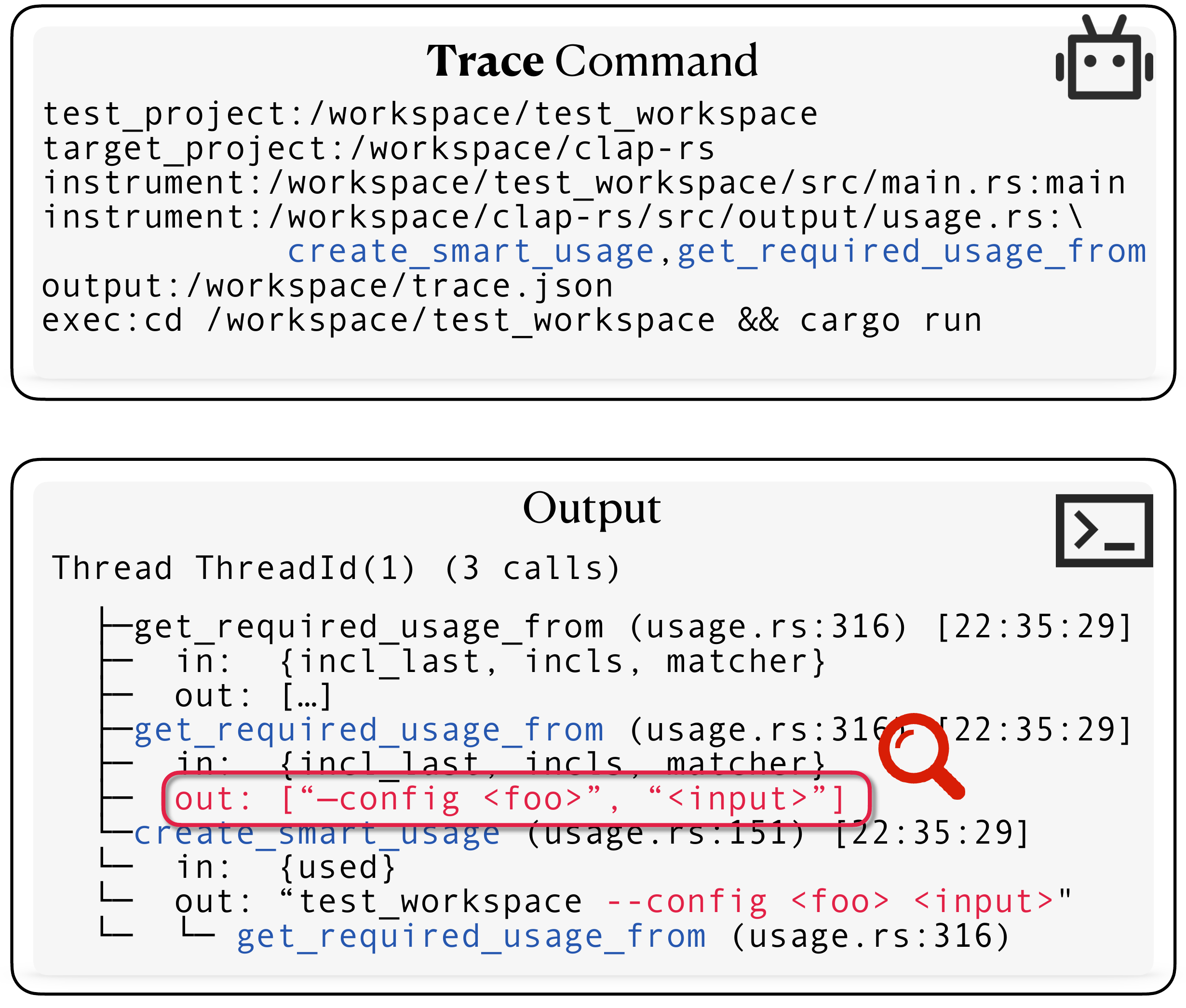}
    \caption{The \texttt{Trace} Command Used in \texttt{clap-rs-2609}~\cite{traj_rustagent_claprs_clap_2609}}
    \label{fig:clap-study}
\end{figure}

To illustrate \rustagent{}'s practical advantages, we analyze its unique resolution of issue \texttt{clap-rs-2609}~\cite{traj_rustagent_claprs_clap_2609}, a task where all baseline agents fail. The issue involves the popular \texttt{clap} library generating an incorrect usage string for arguments with default values. \rustagent{} first executes its automated setup strategy, creating an isolated testing workspace and linking the original project as a local dependency, while other agents consistently fail to reproduce the issue due to the complex build configurations. Subsequently, the agent invokes the \texttt{Trace} command to perform cross-project dynamic analysis on the functions responsible for usage generation as in Figure~\ref{fig:clap-study}. Specifically, \texttt{Trace} automates the entire workflow by: (1) setting up dependencies for both the \texttt{test\_workspace} and the target codebase \texttt{clap-rs}; (2) instrumenting key functions, i.e., \texttt{main} in the test workspace, and \texttt{get\_required\_usage\_from} and \texttt{create\_smart\_usage} in the \texttt{clap-rs} source code, with a tracing macro; (3) executing the test via \texttt{cargo run} and collecting the resulting trace information; and finally, (4) cleaning up all modifications. Interestingly, the trace data precisely expose the root cause: a faulty interaction where one function \texttt{get\_required\_usage\_from} passes an incorrect list of required arguments to \texttt{create\_smart\_usage}. This deep runtime insight enables the agent to correctly deduce a two-part patch modifying both functions. Finally, the agent iteratively validates its solution within the isolated workspace, leveraging the rapid feedback loop to efficiently converge on the correct solution. 

\subsection{Ablation Study}

We further perform an ablation study on 200 randomly selected \rustbench{} tasks, comparing the full \rustagent{} against two variants: \rustagent{}$_{reproduce}$, which disables the \texttt{Trace} command, and \rustagent{}$_{base}$, which skips the isolated testing workspace entirely and directly performs the reproduction in the target codebase.

\begin{table}[t]
    \centering
    \caption{Ablation Study Results}
    \label{tab:ablation-study}
    \setlength\tabcolsep{0.8pt}
    \resizebox{\columnwidth}{!}{
        \begin{tabular}{lccccc}
            \toprule
            \textbf{Configuration} & \textbf{\claudesonnet{}} & \textbf{\gpto{}} & \textbf{\omini{}} & \textbf{\qwen{}} & \textbf{Avg.} \\
            \midrule
            \rustagent{}$_{base}$      & 42          & 12           & 26            & 8  & 22.0 \\
            \rustagent{}$_{reproduce}$ & 47          & 14           & 29            & 10 & 25.0 \\
            \textbf{\rustagent{}}      & \textbf{59} & \textbf{16}  & \textbf{35}  & \textbf{11} & \textbf{30.3} \\
            \bottomrule
        \end{tabular}
    }
\end{table}

Table~\ref{tab:ablation-study} presents the results of our ablation study. First, comparing \rustagent{}$_{base}$ with \rustagent{}$_{reproduce}$, the introduction of the isolated testing workspace alone improves the task resolution performance by resolving 3.0 more tasks on average. 
Furthermore, the performance leap from \rustagent{}$_{reproduce}$ to the full \rustagent{} underscores the value of the cross-project dynamic tracing analysis where the \texttt{Trace} command enables the agent to resolve an additional 5.3 tasks on average. Such results reflect the effectiveness of the issue reproduction and the dynamic tracing analysis of \rustagent{} respectively.

\cmr{To better understand where \texttt{Trace} helps, we also examine two representative issues. In \texttt{bevyengine/bevy-16747}~\cite{traj_rustagent_bevyengine_bevy_16747}, \rustagent{} must extend the \texttt{animated\_field!} macro to support tuple structs (e.g., \texttt{TextColor::0}) while preserving the behavior for named fields. By invoking \texttt{Trace} on the underlying implementation methods, the agent verifies at runtime that calls for named fields still go through \texttt{new\_unchecked}, whereas tuple indices are correctly dispatched to \texttt{new\_tuple\_unchecked}, allowing it to validate the updated macro’s semantics beyond static inspection. In \texttt{tokio-rs/tracing-1017}~\cite{traj_rustagent_tokiors_tracing_1017}, \rustagent{} uses \texttt{Trace} to validate a performance optimization that skips Thread-Local Storage checks when no scoped dispatcher is active. By tracing \texttt{set\_global\_default} and \texttt{with\_default} in the reproduction test, it observes that the global dispatcher is initialized before scoped guards, providing runtime evidence that the conditional bypass does not compromise correctness or thread-safety. These cases concretely illustrate how dynamic tracing contributes to the gains observed in the ablation study.}

\section{Threats to Validity}

\noindent \textbf{Threats to internal validity.} We take several measures to ensure the internal validity of our study. To guarantee reproducibility and mitigate system-level variations, we conduct each experimental run in a dedicated Docker container that encapsulates the exact repository state, toolchain, and dependencies. Furthermore, our \texttt{Trace} command implements a meticulous cleanup mechanism to remove all instrumentation artifacts after its execution, regardless of the outcome, ensuring our analysis introduces no residual side effects to the target codebase.

To mitigate the inherent stochasticity of LLM-based agents, we perform our evaluation across 500 diverse instances in \rustbench{}, ensuring the stability and generalizability of our findings. We set the temperature to 0 to mitigate the threat of randomness to the \rustagent{} evaluation results. Potential bias in our manual validation is mitigated through a strict protocol where three authors independently verify each result before reaching a consensus.

\noindent \textbf{Threats to external validity.} The primary threat to external validity lies in the generalizability of our findings, which is closely tied to our evaluation dataset and the selection of compared agents. Correspondingly, we construct \rustbench{} with 500 real-world, issue-resolving tasks sourced from 34 diverse and popular open-source Rust repositories. Each task is grounded in an actual merged pull request, ensuring the problems are authentic and representative of genuine software engineering challenges. Meanwhile, we select four recent and representative agents that demonstrate state-of-the-art performance on the widely-recognized \swebench{} benchmark, ensuring our analysis reflects the latest landscape of code agents.

\noindent \textbf{Threats to construct validity.} A potential threat to construct validity lies in whether our evaluation metrics accurately capture the multifaceted nature of the issue resolution task. To mitigate this, we adopt the Resolved Rate (Pass@1), a widely-accepted metric for evaluating task completion in code generation benchmarks~\cite{swebench, mswe, swelancer, hu2025assessingadvancingbenchmarksevaluating}. To provide deeper diagnostic insights, we also measure the Reproduction Success Rate (\%)~\cite{beabrwdki,raraacfbr} to assess an agent's ability to replicate the original issues. 

\section{Conclusion}
In this paper, we introduce \rustbench{}, the large-scale, repository-level benchmark for real-world Rust software engineering issues, comprising 500 issue-resolving tasks from diverse and popular repositories. Our comprehensive study on \rustbench{} reveals that the performance of current LLM-based agents is primarily hindered by challenges in repository-wide code comprehension and \modify{issue} reproduction. To address these limitations, we design \rustagent{}, a novel agentic framework integrating an automated testing workspace with a unique cross-project dynamic tracing capability. Our evaluation demonstrates that \rustagent{} achieves the best performance, resolving 28.6\% of the tasks using \claudesonnet{}, i.e., a 34.9\% improvement over the strongest baseline, and uniquely resolving \approachUQResolve{} issues across all adopted LLMs.

\section*{Data Availability}
All study results, evaluation details, and source code of the \rustbench{} and \rustagent{} are presented in the \git{} page~\cite{githubrepo}.

\begin{acks}
This work is partially supported by the National Natural Science Foundation of China (Grant No. 62372220). It is also partially supported by Ant Group Research Fund.
\end{acks}

\newpage
\clearpage
\bibliographystyle{ACM-Reference-Format}
\bibliography{rustbenchref}
\end{document}